\documentclass[aps,pra,twocolumn,groupedaddress,10pt, floatfix]{revtex4-1}
\usepackage{pgfplots} 
\usepackage{tipa}
\pgfplotsset{compat=newest} 
\pgfplotsset{plot coordinates/math parser=false} 
\newlength\figureheight 
\newlength\figurewidth 
\usetikzlibrary{plotmarks}
\usepackage{graphicx}
\usepackage{bm}
\usepackage{amsmath,amssymb}
\usepackage{upgreek}
\usepackage{xspace}
\usepackage{braket}
\usepackage{verbatim}
\usepackage{textcomp}
\usepackage[hidelinks]{hyperref}
\usepackage[load=physical]{siunitx}
\usepackage{epstopdf}
\usepackage{xcolor}
\newcommand{\sref}[2][{}]{\hyperref[#2]{\ref{#2}#1}} 
\DeclareSIUnit\nK{\nano\kelvin}
\DeclareSIUnit\uK{\micro\kelvin}
\DeclareSIUnit\Vcm{\volt\per\centi\meter}
\makeatletter
\g@addto@macro\bfseries{\boldmath}
\makeatother

\definecolor{mc1}{rgb}{0,0.447,0.741}
\definecolor{mc2}{rgb}{0.85,0.325,0.098}

\newcommand{\rubidium}{$^{\text{87}}\rm{Rb}$}

\newcommand{\lc}{$l$}

\usepackage{fancyhdr}
\usepackage{nicefrac}
\usepackage{datetime} 
\renewcommand{\textonehalf}{1/2}
\newcommand{\textthreehalf}{3/2}
\begin{document}
\title{Ultracold chemical reactions of a single Rydberg atom in a dense gas}

\author{Michael~Schlagm\"{u}ller}
\author{Tara Cubel Liebisch}
\author{Felix Engel}
\author{Kathrin~S.~Kleinbach}
\author{Fabian~B\"{o}ttcher}
\author{Udo~Hermann}
\author{Karl M.\ Westphal}
\author{Anita Gaj}
\author{Robert L\"{o}w}
\author{Sebastian Hofferberth}
\author{Tilman Pfau}
\email{t.pfau@physik.uni-stuttgart.de}
\affiliation{5. Physikalisches Institut and Center for Integrated Quantum Science 
and Technology (IQST), Universit\"{a}t Stuttgart, Pfaffenwaldring 57, 70569 Stuttgart, 
Germany}
\author{Jes\'{u}s P\'{e}rez-R\'{i}os}
\author{Chris H.\ Greene}
\affiliation{Department of Physics and Astronomy,
Purdue University, 47907 West Lafayette, IN, USA}
\date{4 May 2016}

\begin{abstract}
Within a dense environment ($\rho \approx 10^{14}$\,atoms/cm$^3$) at ultracold temperatures ($T < \SI{1}{\uK}$), a single atom excited to a Rydberg state acts as a reaction center for surrounding neutral atoms. At these temperatures almost all neutral atoms within the Rydberg orbit are bound to the Rydberg core and interact with the Rydberg atom. We have studied the reaction rate and products for \textit{nS} $^{87}$Rb Rydberg states and we mainly observe a state change of the Rydberg electron to a high orbital angular momentum \textit{l}, with the released energy being converted into kinetic energy of the Rydberg atom. Unexpectedly, the measurements show a threshold behavior at $n\approx 100$ for the inelastic collision time leading to increased lifetimes of the Rydberg state independent of the densities investigated. Even at very high densities ($\rho\approx4.8\times 10^{14}\text{\,cm}^{-3}$), the lifetime of a Rydberg atom exceeds \SI{10}{\us} at $n > 140$ compared to \SI{1}{\us} at $n=90$. In addition, a second observed reaction mechanism, namely Rb$_2^+$ molecule formation, was studied. Both reaction products are equally probable for $n=40$ but the fraction of Rb$_2^+$ created drops to below \SI{10}{\percent} for $n\ge90$.
\end{abstract}


\maketitle

\section{Introduction}
Ultracold Rydberg atoms are studied in increasingly dense environments as they give rise to collective phenomena in the strongly blockaded regime~\cite{Heidemann2007,Dudin2012,Ebert2014,Zeiher2015}, Foerster-type energy transfers~\cite{Guenter2013}, Rydberg-dressed ensembles~\cite{Zeiher2016} or to achieve larger optical depths for quantum optical applications~\cite{Baur2014,Gorniaczyk2014,Tiarks2014}. At increasingly higher densities it becomes more likely that ground state atoms reside within the orbit of the Rydberg-electron and ultralong-range Rydberg molecules~\cite{Greene2000,Bendkowsky2009} can be created with exotic butterfly~\cite{Niederpruem2016} and trilobite~\cite{Li2011} shapes. The largest density achievable at ultracold temperatures are in quantum degenerate gases, where the backaction of a Rydberg atom on the superfluid can be studied~\cite{Balewski2013}. The generation of phonons by immersing a single Rydberg atom may even be used to image the wavefunction of the Rydberg electron~\cite{Karpiuk2015}. For such advanced schemes, it is necessary that the Rydberg atom stays in its original state sufficiently long to have a measurable impact on the surrounding quantum gas. In a previous study it was already shown that the lifetime of Rydberg atoms inside a dense cloud of atoms is reduced~\cite{Balewski2013} but no quantitative study on the origins of the reduced lifetimes have been performed so far. 

\enlargethispage*{0.5cm}

Prior to the present work, inelastic collisions between Rydberg atoms and neutral particles have been investigated but in regimes of energies and densities that are completely different from those of the present work. Chemi-ionization reactions creating Rb$_2^+$ were examined in the existing literature~\cite{Barbier1987,Beigman1995,Kumar1999,Mihajlov2012, Niederpruem2015} leading to the ionization of the Rydberg electron as a consequence of the short-range interaction between the ionic core and the incoming neutral atom~\cite{Miller1970}. Inelastic collisions of highly excited Rydberg atoms in a dense background gas at room temperature were observed in the late 1970's \cite{Gallagher1975,Gallagher1977,Gallagher1978} and early 1980's~\cite{Hugon1982}, and simultaneously the theoretical framework associated with such a process was developed by adopting a semiclassical description of the dynamics, in combination with a constant Rydberg electron-neutral atom scattering length~\cite{Gersten1976,Olson1977,Hickman1978} in applications of the Fermi pseudopotential. This approach describes satisfactorily the overall behavior of the $l$-mixing cross section as a function of the principal quantum number $n$, in a regime where the collision energy between the neutral and the Rydberg atom, $E_{k}$, exceeds the relevant energy scale of the Rydberg transitions $\Delta E_n$, i.e.,\ $E_{k} \gg \Delta E_{n}\sim n^{-3}$, thus justifying the semiclassical treatment. However, for a Rydberg excitation ($n=100$) in an ultracold ($E_{k} \sim 10^{-6} \Delta E_{n}$) and a dense background gas ($\hat\rho = 5.8\times10^{14}$\,cm$^{-3}$), more than 1000 neutral atoms are within the Rydberg orbit. This requires a new theoretical treatment for inelastic Rydberg-neutral collisions to be developed. 
 
\begin{figure}
\begin{center}
\includegraphics[width=0.45\textwidth]{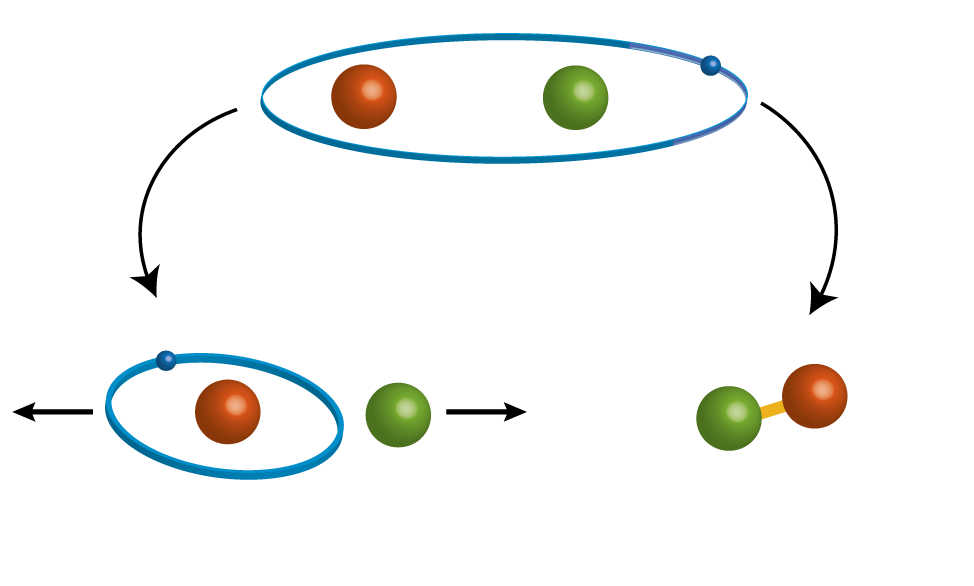}
\begin{picture}(0,0)
\put(-203,100){I}
\put(-35,100){II}
\put(-104,90){Rb}
\put(-175,90){Rb*($n, l=0$)}
\put(-215,10){Rb*($n-4,l>2$)}
\put(-58,10){Rb$_2^+$}
\end{picture}
\caption{\label{fig:Reaction} The two observed reaction channels for a single Rydberg atom in an ultracold and dense environment. The red sphere depicts the Rydberg ionic core and the Rydberg electron is shown in blue, whereas the neutral atoms are shown in green. (I) The Rydberg atom changes its angular momentum and both collision partners share the released energy as kinetic energy leading to an l-mixing reaction channel. (II) A deeply bound Rb$_2^+$ molecular ion is formed through chemi-ionization by an associative ionization reaction.}
\end{center}
\end{figure}
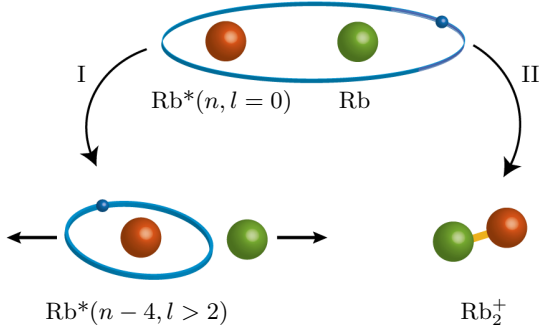

In this work, we study the role of the ionic Rydberg core, the Rydberg electron and the neutral ground-state atom collisions across a large range of principal Rydberg quantum numbers in limiting the collisional lifetime of Rb Rydberg atoms in a dense, ultracold atom cloud. We identify two reaction products, shown in Fig.~\ref{fig:Reaction}, which are explained using a new theoretical framework. A quantum mechanical treatment of the reaction pathways is implemented along with a semiclassical description of the interactions at short internuclear distances, with an explicit treatment of the neutral-Rydberg energy mediated by the Rydberg electron-neutral collisions. Finally, the reaction dynamics and the branching ratio of these ultracold chemical reactions are analyzed.

\begin{figure}
\begin{center}
\includegraphics{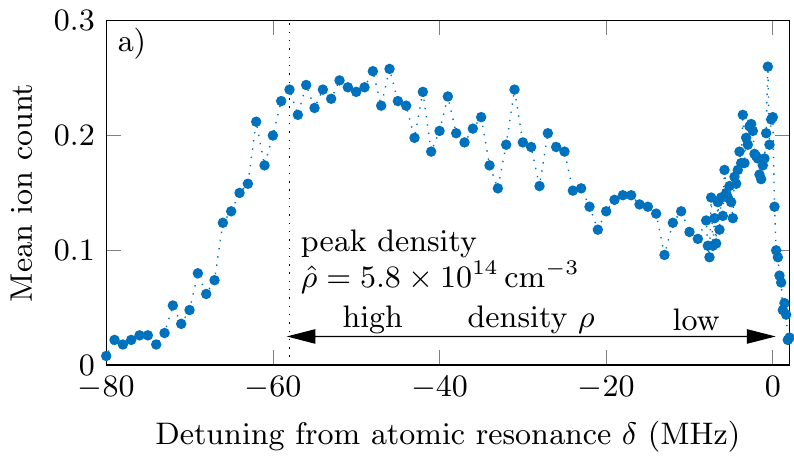}\\\includegraphics{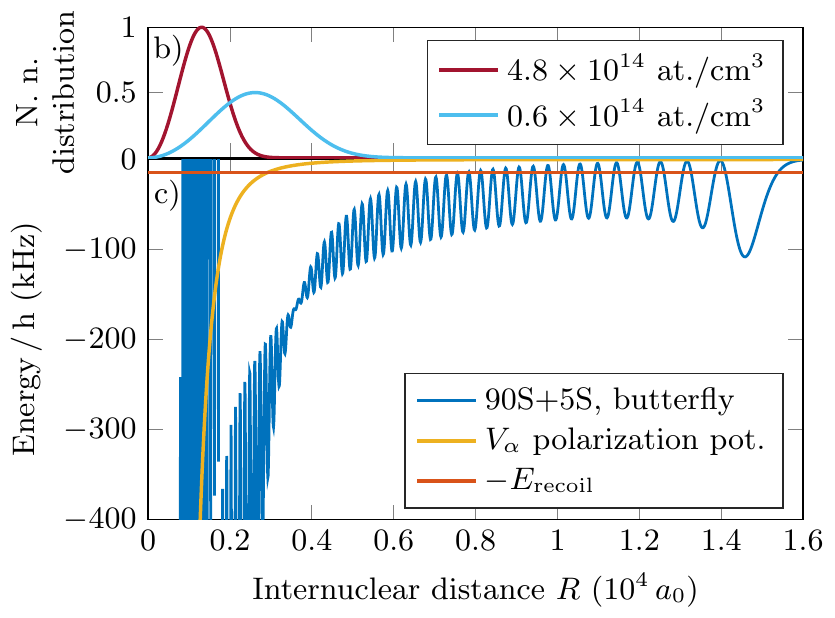}
\caption{\label{fig:idea}(a) Spectrum of a 90\textit{S} Rydberg atom in a BEC. One of the excitation lasers is focused to the center of the BEC with a focal waist of \SI{2.1}{\um}. Therefore, primarily the peak density $\hat \rho=5.8\times10^{14}$\,cm$^{-3}$ of the BEC is addressed with a detuning of approximately \SI{-58}{\MHz}. Each data point is taken with 10 BECs, and each BEC was probed 50 times. During the subsequent experiments within one BEC, the peak density drops from $5.8\times10^{14}$\,cm$^{-3}$ to $4.9\times10^{14}$\,cm$^{-3}$ and approximately $3.5\times10^5$ atoms of the initial $1.7\times 10^6$ of the BEC are in the thermal cloud at the end of the experiment. The thermal atoms surrounding the BEC contribute to the signal at very small red detuning. The laser detuning is related to the local density where the Rydberg atom is excited~\cite{Liebisch2016}. \label{fig:Spectrum90S}(b) Nearest neighbor distribution of the distance between the Rydberg core ($R=\SI{0}{\bohr}$) and a neutral atom for two different densities of the BEC. (c) Pair potential of a 90\textit{S}+5\textit{S} state including the Rydberg electron-neutral scattering interaction and the Rydberg core-neutral polarization potential $V_\alpha$. The state crossing visible at \SI{1700}{\bohr} is the butterfly state~\cite{Greene2000} crossing the $90S+5S$ state. The highest energy available in the system comes from the photon recoil, $E_\text{rec} = h \times \SI{15}{\kHz}$ and therefore most of the particles inside the Rydberg orbit are classically trapped.}
\end{center}
\end{figure}

\section{Experimental setup}

The reaction dynamics are studied in a nearly pure BEC of approximately $\num{1.7e6}$ \rubidium{} atoms in the magnetically trapped spin polarized ground-state $\ket{5S_\text{\textonehalf}\text{,\,}\textit{F}\text{\,=\,2,\,}\textit{m}_{\textit{F}}\text{\,=\,2}}$ produced in a QUIC trap~\cite{Esslinger1998}. The trapping frequencies are $\omega_{r}=2\pi\times\SI{200}{\Hz}$ in the radial and $\omega_{ax} = 2\pi\times\SI{15}{\Hz}$ in the axial direction corresponding to Thomas-Fermi radii~\cite{Pethick2002} of \SI{5.1}{\um} by \SI{68}{\um}. The atom number and trap frequencies give rise to a peak density of $5.8\times10^{14}$\,atoms/cm$^3$ in the BEC. The Rydberg excitation is created by a two-photon excitation scheme, which couples the ground state to the $\ket{\textit{nS}_\text{\textonehalf}\text{,\,}\textit{m}_{\textit{S}}\text{\,=\,1/2}}$ state for principal quantum numbers \textit{n} from 40 to 149, off-resonant via the intermediate state 6\textit{P}$_{\!\text{\textthreehalf}}$. Both excitation lasers are pulsed simultaneously with a repetition rate of \SI{2}{\kHz}, enabling up to 400 experiments within a single BEC at a fixed laser frequency. The excitation probability per shot is kept well below one. The possible single Rydberg atom in the BEC is subsequently ionized by an electric field pulse, with a \SI{200}{\ns} rise time and a variable delay time. The atom number and density of the BEC drops during the repeated measurements but the extracted values were tested to be independent of this atom loss. Further details of the experiment are described in~\cite{Schlagmueller2016,Liebisch2016}.

\section{Neutral atoms bound within a Rydberg orbit}

Under these conditions, a Rydberg atom is excited with, on average, one ($n = 40, \rho = \SI{1e14}{\cm^{-3}}$) or up to tens of thousands ($n = 149, \rho = \SI{5e14}{\cm^{-3}}$) of neutral atom perturbers inside the Rydberg orbit, depending on the principal quantum number $n$ and the density. The neutral atoms inside the Rydberg orbit cause a frequency shift for the excitation of a Rydberg atom and can be used to modify the resonant density region in the BEC~\cite{Liebisch2016}: In a first approximation, the resonance frequency is shifted in relation to the local density, $\rho$, in which the Rydberg atom is excited, due to the Fermi pseudopotential as 
\begin{align}
\label{eq:FermiShift}
\delta(\rho) = \frac{2 \pi \hbar^2 a}{m_e} \rho
\end{align}
with the reduced Planck constant as $\hbar$, the electron mass as $m_e$, and the $s$-wave scattering length as $a$. This leads for triplet scattering in rubidium ($a_{\uparrow\uparrow} = \SI{-15.7}{\bohr}$~\cite{Boettcher2016}) to a spectral line shift of \SI{-10}{\MHz} for $10^{14}$\,atoms/cm$^{\text{3}}$ independent of $n$. The situation is depicted also in Fig.~\ref{fig:Spectrum90S}, showing a 90\textit{S} spectrum of a single Rydberg atom in a BEC for which a laser detuning of \SI{-58}{\MHz} approximately addresses the peak density $\hat \rho=5.8\times10^{14}$\,cm$^{-3}$.

The temperature of the ultracold atoms is approximately \SI{300}{\nK}. The largest energy available in the system comes from the recoil energy resulting from the two-photon excitation, and is $h \times \SI{15}{\kHz}$ ($k_B \times \SI{700}{\nK}$), where $h$ is the Planck constant and $k_B$ the Boltzmann constant. At these low temperatures, many neutral atoms are trapped inside the Rydberg orbit, as shown in Fig.~\sref[b]{fig:idea}, by the Rydberg electron interaction potential and the polarization potential from the ionic core of the Rydberg atom.

The atoms inside the Rydberg orbit approach the Rydberg core after the Rydberg atom is excited, leading to inelastic and reactive processes, namely $l$-mixing collisions and chemi-ionization as depicted in Fig.~\ref{fig:Reaction}. Since the neutral atoms start in an interacting frame, the usual concepts of scattering theory are not well-defined in this case as there are no asymptotic initial states. The figure of merit is therefore the collision time observed instead of a collision rate. This time is much shorter than the radiative lifetimes of Rydberg atoms, making a dense atom cloud an ideal test bed for studying ultracold Rydberg-neutral atom collisions.

\section{Reaction I: Angular momentum changing processes}

\begin{figure}
\begin{center}
\includegraphics{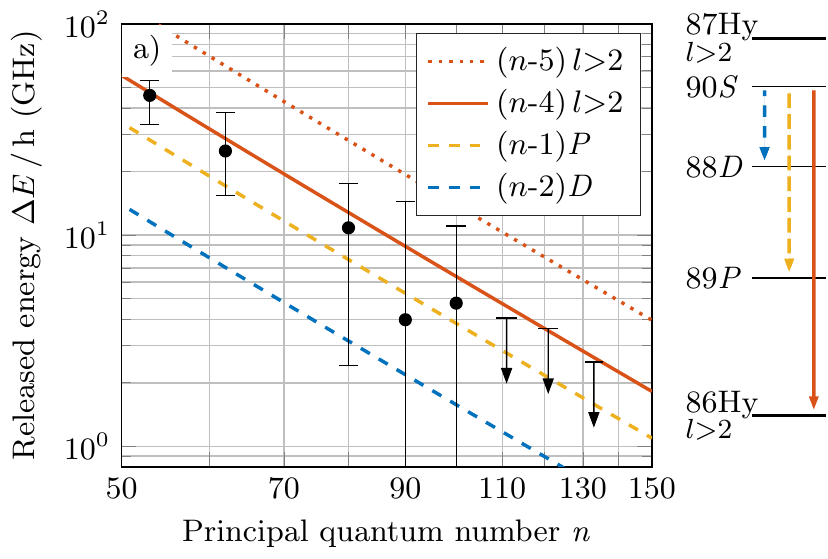}\\\includegraphics{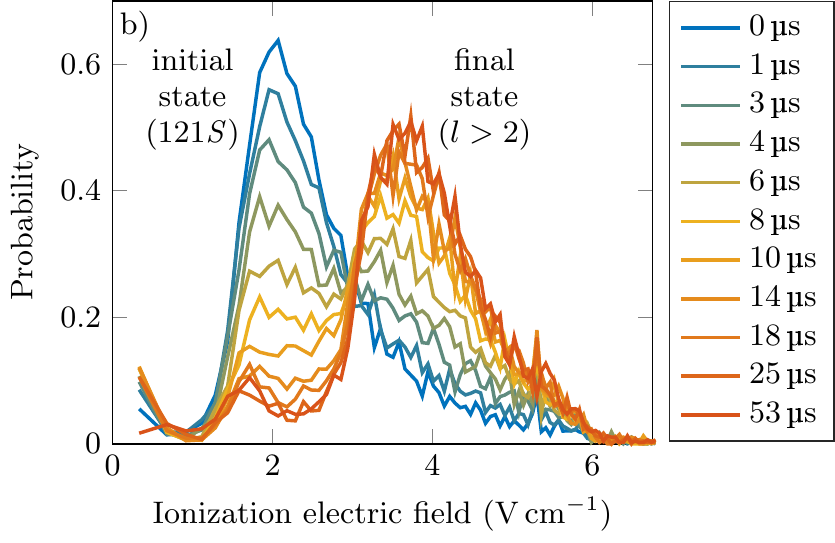}
\caption{\label{fig:nlchange} (a)~Released energy during the \lc{}-changing collision. The black dots indicate the most likely released energy during a collision, whereas the bars indicate the full width at half maximum of the fitted probability distribution of the released energy. The energies observed correspond to a state change to the next lower hydrogenic manifold ($n-4$). For $n\ge110$ the most likely kinetic energy gained is not shown anymore because the released energy is too low to extract a well-defined maximum out of the energy distribution, which is depicted by the arrow. The colors of the released energies are according to the level schema shown on the right, in which Hy denotes the hydrogenic manifold. (b)~State selective ionization analysis of the detected ions for a $121S$ state. With an immediate (blue curve) linear ionization field ramp going to \SI{6}{\Vcm} within \SI{3}{\us} applied after the \SI{500}{\ns} excitation pulse, almost all atoms arrive between 1 and \SI{3}{\Vcm} corresponding to the initially excited $121S$ state. With a delayed ionization, the Rydberg atom undergoes an inelastic collision (red: delay of \SI{53}{\us}) and the required ionization voltage changes by a factor of 2 to 4 corresponding to a high \lc{}-state~\cite{Gallagher1994}.}
\end{center}
\end{figure}

The state-changing collision of the studied $^{87}$Rb Rydberg atom with the neutral atom reagent is an exoergic process, whereby a state-change occurs from the excited \textit{nS} state to the lower-lying hydrogenic manifold. In the case of rubidium, with a quantum defect of the \textit{S} state ($\delta_S = 3.13$~\cite{Mack2011}), this is the $(n-4)$ hydrogenic manifold. We can measure the change in kinetic energy of the Rydberg atoms undergoing this collision process by monitoring the arrival time of the ionized Rydberg atoms on the microchannel plate detector. For this measurement the Rydberg atoms are ionized after a delay time, e.g.\ $\tau_\text{delay} = \SI{100}{\us}$, which is significantly longer than the collisional lifetime $\tau_\text{coll}$ of the Rydberg states. After the \lc{}-changing collision, the Rydberg atom and the neutral atom fly apart in opposite directions, due to momentum conservation and their equal mass, with respect to the excitation location. Once the ionization pulse is applied at time $\tau_\text{delay}$, the Rydberg atom accelerates towards the microchannel plate detector. The varying spatial positions of the Rydberg atom at the time of ionization are detected as a spread in ion arrival time. Using ion trajectory simulations, we calculate the expected ion arrival times based on the distance \textit{d} from the center where the Rydberg atoms are excited. This mapping of initial position to ion arrival time determines the average kinetic energy of the detected Rydberg atoms. The total energy $\Delta E$ released in this reaction is twice the energy of the detected Rydberg atom $E_\text{kin}$, because the energy released during the collision must be shared between both collision partners:
\begin{equation}
\label{eq:EnergyRelease}
\Delta E = 2 E_\text{kin} = m_{\text{Rb}} \left(\frac{d}{\tau_\text{delay} - \tau_\text{coll}}\right)^2.
\end{equation}
Further details for this method are given in appendix~\ref{a:EnergyRelease}.

Fig.~\sref[a]{fig:nlchange} plots the released energy, gained during the inelastic collision, versus the principal quantum number. The released energy $\Delta E$ corresponds for low principal quantum numbers well to the change in potential energy from the excited $nS$ Rydberg state to the hydrogenic manifold below, which is the $(n-4)$ manifold because of the quantum defect of Rb with $\delta_S = 3.13$. The energy $\Delta E$ can be calculated according to the Rydberg formula~\cite{Ritz1908}
\begin{align}
\label{eq:Rydberg}
\Delta E &= Ryd_\text{Rb} \left( \frac{1}{(n_1^*)^2} - \frac{1}{(n_2^*)^2}\right) \\
&= Ryd_\text{Rb} \left( \frac{1}{(n-\delta_S)^2} - \frac{1}{(n-4)^2}\right)
\end{align}
with the effective principal quantum number $n^*$ taking into account the quantum defect, and the modified Rydberg constant $Ryd_\text{Rb}$ for $^{87}$Rb. Due to the decreasing energy level spacing, the released energy $\Delta E$ decreases with higher principal quantum numbers. Furthermore, the change in kinetic energy as confirmed by the measurement shown in Fig.~\sref[a]{fig:nlchange} strongly indicates that we do not couple for any investigated $n$ to the ($n-5$) hydrogenic manifold or any levels with larger energy separations. 

The threshold ionization electric field of the Rydberg atoms can be exploited to examine the angular momentum change of the Rydberg atom. At a slew rate of the ionization electric field of $\SI{2}{\Vcm \per \us}$, Rydberg atoms with $l>2$ ionize diabatically at up to four times the adiabatic (classical) ionization threshold field ($1/16n^{*4}$)~\cite{Gallagher1994, Walz-Flannigan2004,Guertler2004}. Therefore, the higher electric field required to ionize the Rydberg atom, which is visible in Fig.~\sref[b]{fig:nlchange}, shows that the final state must be a high \lc{}-state. 

Based on the measured released energy and the ionization threshold change combined, the most likely populated final state after the inelastic collision of this reaction channel is the $(n-4)$ hydrogenic manifold below the excited $nS$ state. The energy release restricts the possible state change to one effective principal quantum number below the initial state. For low $n$, the most probable released energy is in accordance to a state change to the next lower lying manifold. For high principal quantum numbers, for which the energy resolution is not precise enough to determine the target state, the state selective ionization measurements show that a high \lc{}-state is populated.
 
\begin{figure}
\begin{center}
\includegraphics{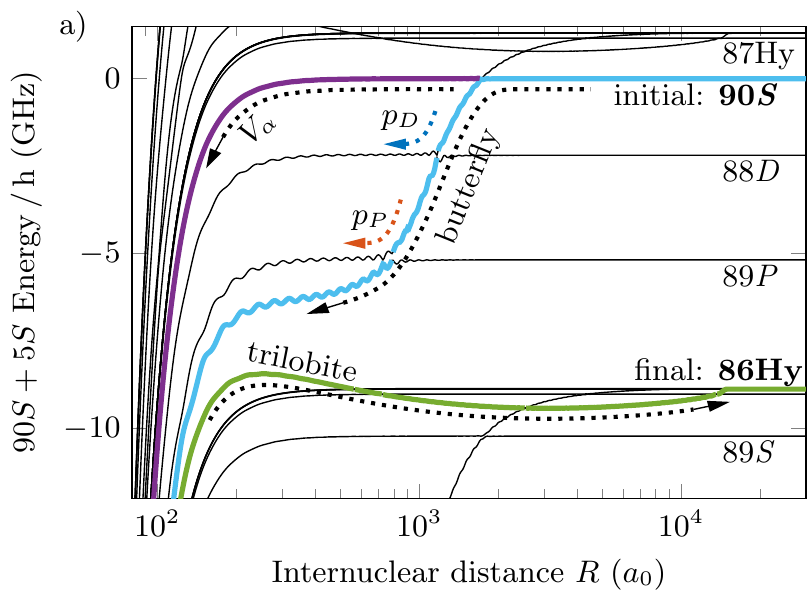}\\\includegraphics{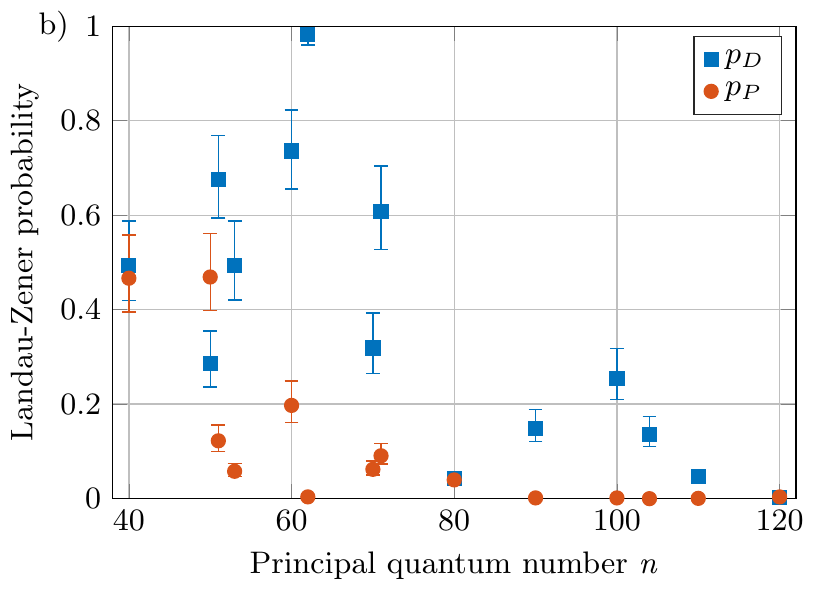}
\caption{\label{fig:PEC} (a) Potential energy landscape of the $90S$ state including the interaction between the Rydberg electron and a neutral atom including the polarization potential $V_\alpha$ of the Rydberg ionic core. At $R\approx\SI{1700}{\bohr}$, the $90S$ and the neighboring butterfly state couples and causes an avoided crossing. From a classical viewpoint, neutral atoms initially at shorter distances than the crossing will follow the $S$ potential until the ion neutral interaction takes over (violet). At larger distances, the probability to adiabatically follow the butterfly state at the crossing of the $nS$ potential with the butterfly state is almost one, as the neutral atom approaches the Rydberg ionic core. In both cases a short range coupling at $R < \SI{200}{\bohr}$ can lead to a state change to the trilobite state (green), which for large internuclear separations turns into a high \lc{}-state of the next lower lying hydrogen manifold (Hy). (b) Theoretical adiabatic Landau-Zener probabilities at the butterfly crossing with the \textit{D} and \textit{P} states, respectively, as functions of the principal quantum number $n$. The error bars have been calculated as it is explained in the appendix~\ref{a:LZ}.}
\end{center}
\end{figure}

\begin{figure}
\begin{center}
\includegraphics{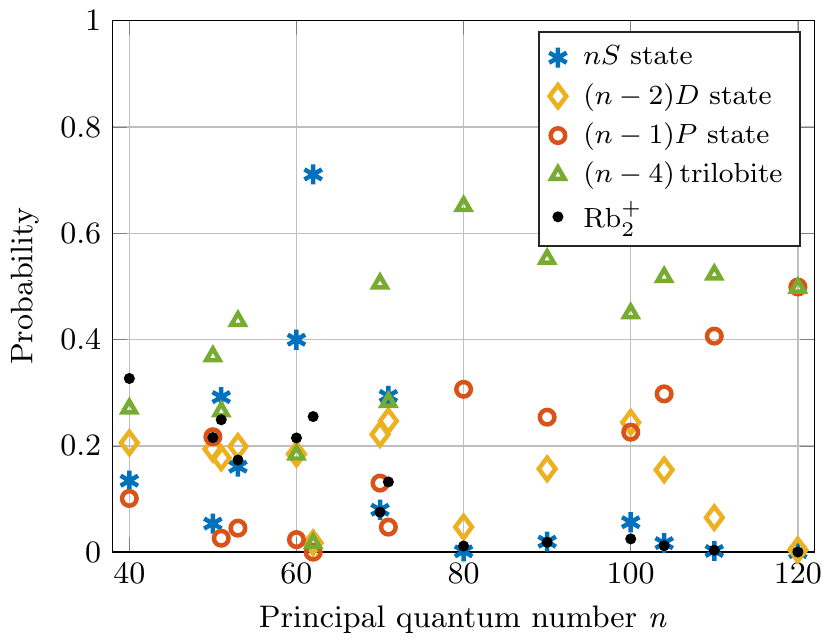}
\caption{\label{Coupling} Theoretical results for the different pathways. Probability to find the Rydberg atom in different states after an ion-neutral collision: $S$~state (blue), $P$~state (red), $D$~state (orange) and in the $(n-4)$~trilobite state (green). The Rb$_2^+$ reaction product will be discussed in the section~\ref{s:Rb2}.}
\end{center}
\end{figure}

The Rydberg-neutral collision is determined by the molecular potential energy landscape. In particular for distances below $r < \SI{1800}{\bohr}$, it is known that the electron-neutral $p$-wave shape resonance at \SI{0.03}{\eV} significantly affects this landscape~\cite{Greene2000, Chibisov2002, Schlagmueller2016}. We thus calculate the molecular potentials by means of Green's function approach~\cite{Hamilton2002} including energy-dependent $s$-wave and $p$-wave triplet scattering phase-shifts for e$^{-}$-Rb collisions. As an example, the potential energy curve (PEC) associated with the 90S state is shown in panel (a) of Fig.~\ref{fig:PEC}. 
To study the collision dynamics, we follow the trajectory of an initial pair state $nS + 5S$ when the two collision partners approach each other. At each avoided crossing in the PECs, we determine the probability for the initial state to follow a specific path by applying the Landau-Zener formula~\cite{Landau1977} including the relative velocity of the two partners at each crossing. To begin with, the probabilities of ending up in either the $(n-2)D$ or the $(n-1)P$ state, is denoted as $p_{D}$ and $p_{P}$, respectively are shown in panel (b) of Fig.~\ref{fig:PEC}. 

Landau-Zener probabilities strongly depend on the kinetic energy at the crossing point, which in the present approach is dominated by the energy difference between the initial \textit{nS} state and the final state reached via the butterfly state pathway. At our very low temperatures, the Rydberg level splitting by far dominates over the initial kinetic energy of the atoms, which was not the case for experiments at room temperature~\cite{Gallagher1975,Gallagher1977,Gallagher1978,Hugon1982}. In panel (b) of Fig.~\ref{fig:PEC}, aside from the noticeable variance in principal quantum number, one observes an overall trend: at low principal quantum numbers $p_{D}$ and $p_{P}$ are systematically larger in comparison with $n \ge 80$. This translates into a larger population of the butterfly state at high principal quantum numbers. The varying probability in $p_{P}$ and $p_{D}$ is related with the derivative of the $(n-1)P$ and $(n-2)D$ Rydberg wave function, respectively at the $p$-wave crossing point. A more detailed explanation of the calculation of the Landau-Zener probabilities is given in appendix~\ref{a:LZ}.

For Rydberg-neutral distance $R \sim \SI{300}{\bohr}$ the butterfly and $(n-1)P$ states are energetically close to the trilobite state coming from the $(n-4)$ hydrogenic manifold, as shown in Fig.~\sref[a]{fig:PEC}, leading to a coupling between these states. The non-adiabatic couplings of the involved electronic states have been quantified by means of the $\mathcal{P}$-matrix $\mathcal{P}_{ij}=\langle \phi_{i} (R)|\frac{ \partial}{\partial R} |\phi_{j}(R) \rangle$ \cite{Clark1979}, where the adiabatic states $|\phi_{i}(R)\rangle$ and $|\phi_{j}(R)\rangle$ have been obtained by diagonalizing the Hamiltonian by means of a truncation of the Hilbert space~\cite{Schlagmueller2016} instead of the Green's function method. As it is shown in appendix~\ref{a:pmatrix}, the $(n-2)D$ state has a negligible coupling with respect to the trilobite state, and hence every atom ending up in the $(n-2)D$ state after passing through the butterfly region will reach the short-range region dominated by chemical forces. However, atoms in the $(n-1)P$ state or butterfly state after the butterfly region experience a considerable coupling to the trilobite state correlated with the $(n-4)$ hydrogenic manifold. Even though atoms will have a chance to reach the short-range interaction region in the trilobite state, Rydberg atoms in high angular momentum states, such as those forming trilobite states, show an extremely narrow autoionization resonance width~\cite{Gallagher1994}. This translates into a very small probability of chemi-ionization reactions. Figure~\ref{Coupling} displays the probability to find the Rydberg atom in a given state after a collision with the perturber. These probabilities have been calculated by analyzing all the available reaction pathways as well as accounting for the coupling among them.

At short Rydberg-neutral distances the ion-neutral polarizability attraction dominates the electron-neutral interaction. This leads to associative ionization, an autoionization process, to be elucidated in the next section. However, the characterization of all the relevant couplings associated with the adiabatic Rydberg-neutral states allows us to calculate the probability to find the Rydberg atom in a particular state after a short-range collision, i.e., when the neutral leaves the butterfly region on its way back to the long-range region.  Due to the fast timescales of the short-range physics we can only resolve the final states in the long-range region, in our experimental setup. The results for the relevant pathway probabilities are shown in Fig.~\ref{Coupling}, where the associative ionization probability of reaction given by Eq.~(\ref{eq:ai}) has been included. The associative ionization probability leads to the second reaction channel, namely the Rb$_2^+$ formation. Even though the model predicts the right fraction for the molecule formation, which will be discussed in the next section, there is a discrepancy for the $l$-changing collision at $n=120$. The theory predicts a probability of \SI{50}{\percent} to end up in the $P$~state whereas the measurement (Fig.~\sref[b]{fig:nlchange}) shows a smaller fraction ($<\SI{20}{\percent}$) for the atoms which ionize at the lower electric field ($l\le 2$).

For the theoretical results shown in Fig.~\ref{Coupling}, the contribution of neutral atoms at distances closer to the Rydberg atom than the distance of the butterfly state crossing the $nS$ state has been neglected. Certainly, these atoms will increase the probability to have associative ionization reactions since $S$ states at small internuclear distances are easily ionized. Moreover, $S$ states do not show any coupling with trilobite states.

\section{Reaction II: Rb$_2^+$ formation}
\label{s:Rb2}
\begin{figure}
\begin{center}
\includegraphics{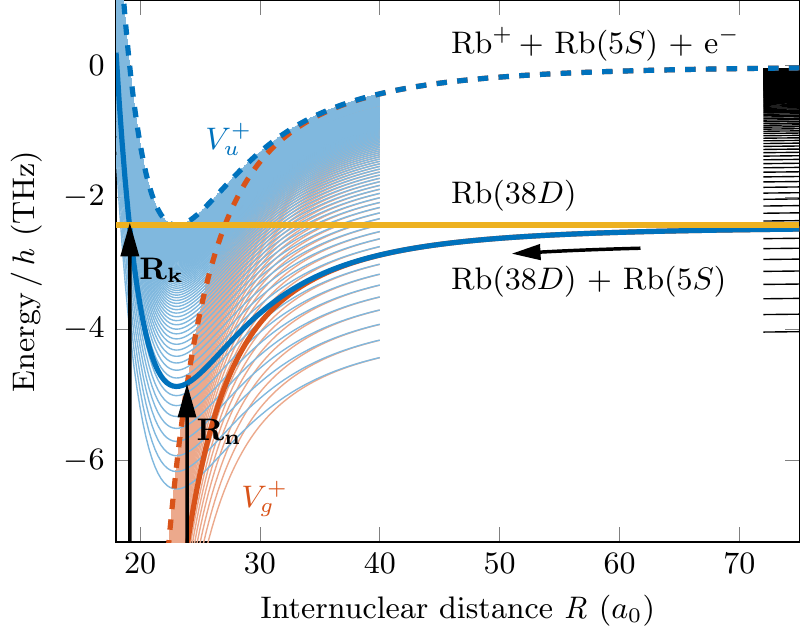}
\caption{\label{fig:Rb2pot} Potential energy curves relevant to the formation of Rb$_2^+$ for Rydberg states from $30D$ to the ionization threshold. For each Rydberg state there are two PEC attached, namely $V_{u}^{+}(R)$ showing a primarily repulsive nature whereas the other labeled $V_{g}^{+}(R)$ displays a far more attractive nature that supports many bound states. The dark blue line represents the potential curve $V_{u}^{+}(R)$ which correlates with the $38D$ state, and the crossing point of this potential with the $V_{g}^{+}$ PEC correlates with the onset where autoionization can begin into the channel Rb$_{2}^++\text{e}^-$ (red dotted line). This crossing point $R_{n}$ represents outer boundary of the associative ionization region. The inner boundary of this autoionization region ends at the inner turning point $R_{k}$, where the collision energy (solid orange line) is equal to the potential.}
\end{center}
\end{figure}

A highly excited atom colliding with a ground state atom can experience a chemi-ionization process~\cite{Mihajlov2012}, which is the ionization of the excited atom as a consequence of the short-range interaction between the ionic Rydberg core of the excited atom and the neutral atom. This leads to associative ionization (AI), which in the case at hand is identified as
\begin{equation}
\text{Rb}^{*}+\text{Rb} \rightarrow \text{Rb}_{2}^{+} +\text{e}^{-}.
\end{equation}

The previous sections demonstrate that a neutral atom interacting with a Rydberg atom transits through a multi-pathway Rydberg energy landscape (see panel (a) of Fig.~\ref{fig:PEC}). At very short internuclear distances $R \lesssim \SI{40}{\bohr}$, however, the electronic cloud of the Rydberg ionic core starts to overlap with the electronic cloud of the valence electron of the neutral atom, leading to a stronger interaction between the neutral ground state atom and the ionic core. This short-range interaction translates into two PECs at short range, which correlate at long range with the Rb$^{+}$\,+\,Rb atomic asymptote. The ground state PEC of the molecular ion is a $^{2}\Sigma_{g}$ electronic state, labeled as $V_g^+$ for simplicity, which supports a large number of bound states, i.e., Rb$_{2}^{+}$, whereas the second PEC is a mostly repulsive electronic state $^{2}\Sigma_{u}$ (weakly attractive between \SI{20}{\bohr} and \SI{\sim 70}{\bohr}), labeled as $V_{u}^+$. 

Let us assume that our initial Rydberg state, an $nS$ state, follows the butterfly state (see Fig.~\sref[a]{fig:PEC}) until the next state crossing and adiabatically changes into the $nD$ state, which is the state it enters the short range region as depicted in Fig.~\ref{fig:Rb2pot}. Attached to this Rydberg state are the $V_{g}^+$ and $V_{u}^+$ electronic states that are degenerate in energy for large internuclear distances. The probability to transfer into either of the two available ($g$ and $u$) states is approximately the same ($p=1/2$) as $R$ decreases through the region and exchange interaction becomes important. Further, we notice that the $V_{u}^+$ state attached to the $nD$ state shows a crossing point at $R_{n}$ with the $V_{g}^+$ potential that correlates with the Rb$^{+}$ + Rb asymptote. Thus, for $R < R_{n}$, $V_{u}^+$ (thick blue curve in Fig.~\ref{fig:Rb2pot}) can be interpreted as the lower boundary for an ionization continuum of potential energy curves $V_{g}^+$ that correlate with the state Rb$_{2}^{+}$\,+\,e$^{-}$. On the other hand, $R_{n}$ also denotes the internuclear distance where the energy splitting between $V^+_{g}$ and $V^+_{u}$ is equal to the Rydberg binding energy in the ungerade bound channel, i.e., $1/2(n-\delta_D)^{2}$, and hence for $R < R_{n}$ it is possible for the Rydberg electron to autoionize, which then forms Rb$_{2}^{+}$ in the state $V^+_{g}$, leading to the product rovibrational state of the AI reaction. The reaction probability for this reaction is given by~\cite{Miller1970}

\begin{equation}
\label{eq:ai}
p_\text{AI}=p\left[1-\exp{\left(-2\int_{R_k}^{R_{n}}\frac{W(R)dR}{\hbar v(R)}\right)}\right],
\end{equation}

\noindent
where $v(R)=\sqrt{2(E_{k}-V_{u}(R))/\mu}$ is the radial velocity for the Rydberg-neutral collision for an $s$-wave atom-atom collisions. $E_{k}$ denotes the collision kinetic energy and $\mu$ is the reduced mass for the colliding partners. $R_{k}$ stands for the inner classical turning point associated with the collision energy $E_{k}$ and $W(R)$ represents the width of the autoionization resonances which are in general proportional to $1/n^{*3}$. 

\begin{figure}
\begin{center}
\includegraphics{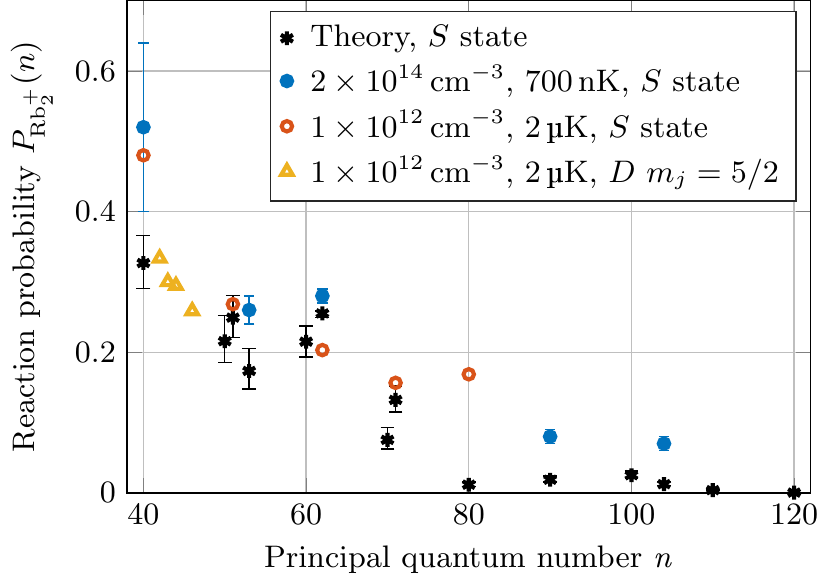}
\caption{\label{fig:Rb2Fraction} The probability $P_{\!\text{Rb}2^+}(n)$ to create deeply bound Rb$_2^+$ molecules is shown, calculated from the measured fraction of ionization events that produce Rb$_2^+$ $P_{\!\text{Rb}2^+}(n) = N_{\text{Rb}2^+} / (N_{\text{Rb}^+} + N_{\text{Rb}2^+})$. The delay time of the ionization in this experiment is chosen such that almost all atoms experienced an inelastic collision. The blue error bars show the experimental uncertainty due to the finite electric field within the science chamber that drags away the molecular ions created. As a comparison, two data sets ($S$~\cite{Gaj2014} and $D, m_j=5/2$~\cite{Krupp2014}) from measurements of a thermal cloud ($T=\SI{2}{\uK}$) with a peak density of only $10^{12}$\,cm$^{-3}$ are added. The black dots represent the theoretical values for the probability of reaction by means of Eq.~(\ref{eq:ai}) and using the corresponding Landau-Zener probabilities. The error bars have been calculated from the errors shown in panel (b) of Fig.~\ref{fig:PEC}}
\end{center}
\end{figure}

Experimentally, the state selective ionization measurement, plotted in 
panel (b) of Fig.~\ref{fig:nlchange}, shows that most of the excited Rydberg atoms undergo an inelastic collision if the delay time before the ionization is long enough. Therefore, the branching ratio between the two reaction channels can be calculated by measuring the fraction of Rb$_2^+$ molecules following an inelastic collision compared to the total signal arriving at the detector. Figure~\ref{fig:Rb2Fraction} shows the resulting reaction probability associated with the formation of Rb$_2^+$. The same figure also shows our theoretical predictions based on Eq.~(\ref{eq:ai}), including the Landau-Zener probabilities shown in panel (b) of Fig.~\ref{fig:PEC}. This calculation utilizes the appropriate collision energy, which turns out to be dominated by the energy difference between adjacent Rydberg states through the butterfly state, as it was indicated in the preceding section. For the purposes of this estimate, the width function has been approximated by the formula $W(R)=0.9/n^{*3}$ in Eq.~(\ref{eq:ai}), which is of the same order as the theoretical limit of $2/\pi$ predicted by two-channel quantum defect theory with one closed Rydberg channel of autoionizing resonances, and this evidently yields overall agreement with our experimental data.

The branching ratio of the reaction products, as shown in Fig.~\ref{fig:Rb2Fraction}, strongly depends on the principal quantum number of the Rydberg atom, but not on the excitation laser detuning, which is related to the atom density, in which the Rydberg atom is excited. For $40S$, the Rb$_2^+$ formation is as likely as an \lc{}-changing collision. The fraction of deeply bound Rb$_2^+$ molecules drops for higher principal quantum numbers to below \SI{10}{\percent} above $n=90$ for the $S$ state. 
The theoretical results following Eq.~(\ref{eq:ai}) show the same trend as the experimental data in this range of principal quantum numbers, however the theoretical predictions are systematically below the observed experimental data at high principal quantum numbers. The reason for this discrepancy might conceivably be associated with our relatively crude utilization of the Landau-Zener approximation for the avoided crossings involving highly excited Rydberg states, since in reality the highly oscillating nature of the Rydberg state wave functions can result in overlapping avoided crossings.

The branching ratio of the low density measurements was extracted out of spectra measured in a thermal cloud at a temperature of $T=\SI{2}{\uK}$~\cite{Gaj2014, Krupp2014}. The excitation laser pulse duration was \SI{50}{\us} and the ionization was applied directly after the excitation. The data points for the $S$ state agree with the high density BEC data, which means that most of the Rydberg atoms went through the chemical reaction during the long excitation time, under the assumption that the branching ratio is independent of the density.

Finally, the molecular ions formed are deeply bound, which was tested by comparing the number of molecules detected for the $62S$ state with an electric field ionization of \SI{180}{\Vcm} and \SI{860}{\Vcm}, which are respectively 7 times and 32 times the classical ionization threshold of the $62S$ state. The number of molecules measured was not lower for higher electric fields, which means that the molecular bond was not disturbed with application of high ionization fields. The same behavior was observed for the $90S$ state (classical ionization field of \SI{5.6}{\Vcm}) with an applied ionization field of \SI{13}{\Vcm} and \SI{180}{\Vcm}. In addition, it was tested that the signal with a field ionization well below the ionization threshold stays the same, which means that the molecules have autoionized during the process. It was found that the effect of the photoionization from the \SI{1020}{\nm} laser is negligible. Measurements in which the \SI{1020}{\nm} laser was kept on after the excitation pulse showed that the number of molecules detected was only approximately \SI{5}{\percent} higher compared to the normal case in which both lasers are turned off simultaneously.

\section{Collisional time for the observed reactions}

\begin{figure}
\begin{center}
\includegraphics{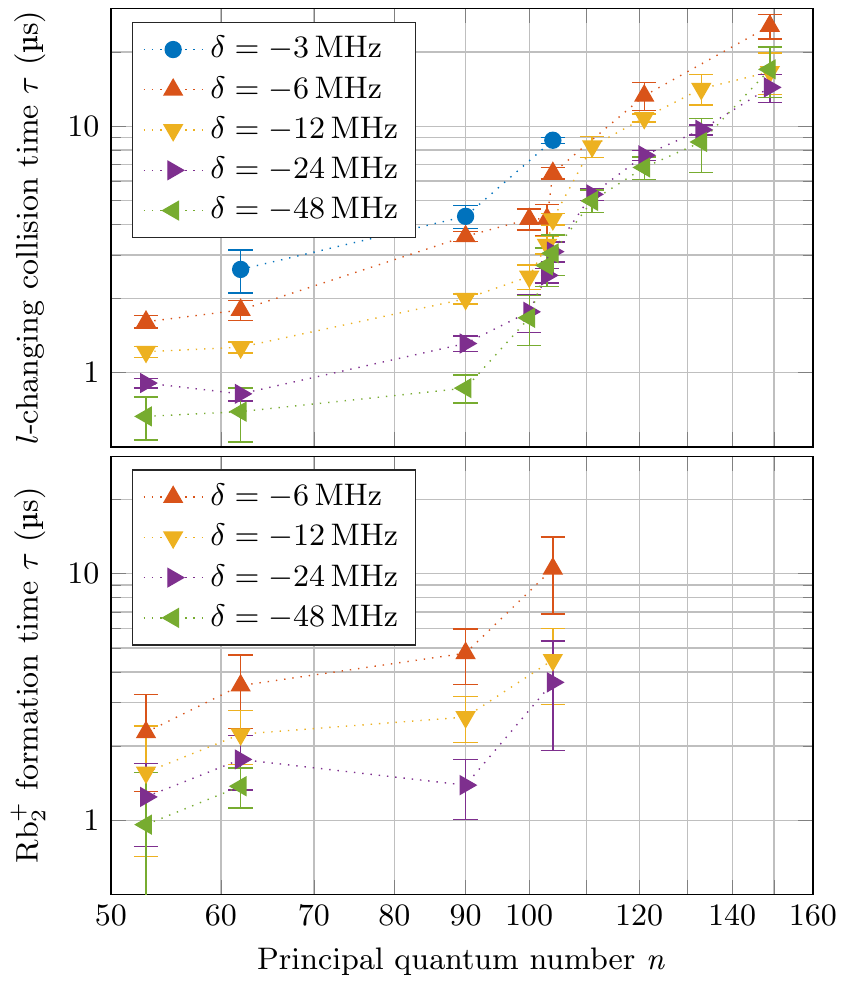}
\caption{\label{fig:taus} Extracted inelastic collision times for the \lc{}-changing collision (top) and the formation of Rb$_2^+$ (bottom). The inelastic collision time $\tau$ of the reactions was extracted out of the measured signal of the reaction products $I(t)$ by fitting $I(t) \propto 1-\exp(-t/\tau)$ with the error bars indicating the confidence interval of the fit taking into account multiple measurements at each data point. Unexpectedly, at around a principal quantum number of $90 < n < 110$, a threshold is reached leading to significantly higher lifetimes of the Rydberg state at higher principal quantum numbers independent of laser detuning and thus the density. The Rb$_2^+$ formation time could not be measured for high $n$ because the total number of molecules measured was too low at these principal quantum numbers.}
\end{center}
\end{figure}

The present experimental approach allows us to measure the amount of products formed as a result of ultracold chemical reactions, as it has been shown. However, the key advantage of the density regime utilized in this study, is the possibility to observe chemical reaction in the time domain, allowing a very near characterization of the two reactions at hand: $l$-mixing collisions and chemi-ionization. The time scale on which this state-changing collision occurs could in principle be examined with the state selective ionization measurement shown in Fig.~\sref[b]{fig:nlchange} by evaluating the time evolution of the measured signal in the initial state and final state. To achieve a better time resolution, however, a two-step ionization was performed with rise times for the electric field of \SI{200}{\ns}. The electric field of the first ionization pulse was chosen such that it ionizes mainly the initial state, but not the final state (e.g.~\SI{2.8}{\Vcm} for $121S$ in Fig.~\sref[b]{fig:nlchange}). A subsequent second ionization pulse, at least four times higher than the classical ionization field, was applied to ionize the remaining Rydberg atoms. All detected atoms of the second ionization pulse must have undergone an \lc{}-changing collision before the first ionization pulse. This fraction is extracted from the data and its evolution with delay time is used to determine the lifetime of the initial Rydberg state, shown in Fig.~\ref{fig:taus}, under the influence of the inelastic collisions. An example of how the collision time is extracted is shown in appendix~\ref{a:Tau}.

\begin{figure}
\begin{center}
\includegraphics{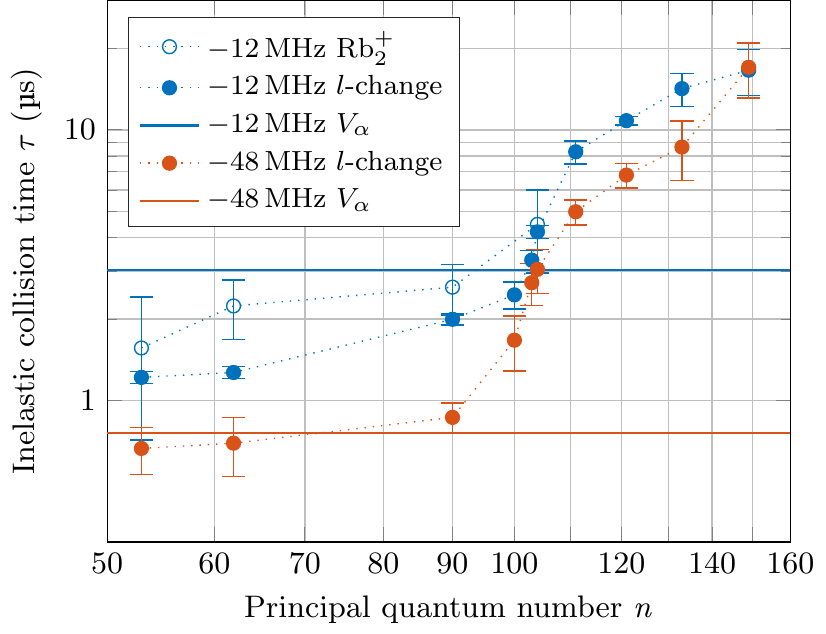}
\caption{\label{fig:tauSim} 
Comparison between the inelastic collision time (see also Fig.~\ref{fig:taus}) and the collision time of a molecular dynamics simulation of an ion and a neutral atom with the polarization potential $V_\alpha$ only. The density for the simulation was chosen such that the density is according to Fermi's density shift (Eq.~\ref{eq:FermiShift}) with $\SI{-10}{\MHz} / (10^{14}$\,cm$^{-3}$). The classical simulation with point-like particles reproduces the order of magnitude for the inelastic collision time at low principal quantum numbers. On the one hand, the full quantum treatment including the electron must accelerate this process for lower $n$, which is very likely due to the butterfly state crossing the $nS$ state. On the other hand, it must slow down the inelastic collision above the threshold region $90< n < 110$ so that collisional lifetimes above \SI{10}{\us} are possible even at the peak density of the BEC.}
\end{center}
\end{figure}

Quite unexpectedly we observe a threshold behavior in the collisional lifetime in the region of $90 < n < 110$. The overall shape of the curves does not depend on the laser detuning and therefore the density, which only causes an overall offset factor to the collision time. The unexpected long lifetime for $n > 110$ is therefore solely a state-dependent effect. The developed theoretical framework to explain the two observed reactions is applicable once the neutral atom crosses the butterfly region. The neutral atom is accelerated due to the butterfly state, leading to collisional times $\SI{\sim10}{\ns}$ for both reactions. Therefore, we conclude that the observed collisional times are related with the physics beyond the butterfly region, which is fully dominated by the Rydberg electron colliding with neutral atoms. On the other hand, taking into account the large number of perturbers within the Rydberg orbit under the present experimental conditions, a molecular dynamic simulation assuming a microcanonical ensemble and including only the ion-neutral interaction was performed, as shown in Fig.~\ref{fig:tauSim}. To simplify the potential, only the attractive force due to the polarizability of the neutral atoms is taken into account in the simulation: The Rydberg core can be treated as an ion for the interaction with the neutral atoms within the Rydberg orbit, because the Rydberg electron does not shield the charge of the core within this region. The Rydberg core and the polarizability $\alpha=\SI{2.6956e-39}{C^2m^2 \per \joule}$~\cite{Holmgren2010} of the neutral rubidium atoms then lead to the polarization potential
\begin{align}
V_\alpha(R) = - \frac{1}{(4\pi\epsilon_0)} \frac{\alpha e^2}{2R^4}
\end{align}
at an internuclear distance $R$, with the vacuum permittivity $\epsilon_0$ and the elementary charge $e$.

The simulated collision time of this system is too slow to explain the observed reaction times for low quantum numbers as shown in Fig.~\ref{fig:tauSim}. In addition, this kind of interaction does not depend on the principal quantum number and therefore cannot reproduce the observed threshold behavior. This is a strong indication that the Rydberg electron plays a key role in the collision dynamics, as one expected based on investigations of the reaction pathways, which was also found to be the case increasing the scattering cross section at similar temperatures~\cite{Niederpruem2015}. The Rydberg electron interacting with the neutral atom can accelerate the collision process of the inelastic collision especially due to the anti-crossing of the butterfly state with the excited $nS$ state, as shown in panel (a) of Fig.~\ref{fig:PEC}, which is the case for lower quantum numbers $n < 90$ with collision times of a few \si{\us}. For higher principal quantum numbers, the collision time exceeds \SI{10}{\us} for $n\ge140$. Here we suspect that a quantum mechanical effect sets in which prevents the neutral atoms for colliding inelastically with the Rydberg atom. A possibility for this could be quantum reflection at the steep slope of the butterfly crossing the initial $nS$ state as depicted in Fig.~\sref[a]{fig:PEC}, which is already known as a mechanism to form bound states in rubidium~\cite{Bendkowsky2010} and depends on the principal quantum number. On the other hand, very recently a similar scenario has been explored finding deviations from the expected additive nature of the many-body landscape for the interaction between Rydberg atoms and neutral atoms~\cite{Eiles2016}, which may be also relevant to understand the measured collisional times.

In Fig.~\ref{fig:tauSim} the collision time of the two different reaction channels, shown in Fig.~\ref{fig:taus}, are compared. The Rb$_2^+$ formation time can be up to two times the \lc{}-changing collision time. It is unlikely that the Rb$_2^+$ molecules are created as a subsequent collision after an \lc{}-changing collision, because for the kinetic energy gained during the collision, the Rydberg atom leaves the dense BEC region within a microsecond and the scattering cross section is reduced, due to the higher velocity (e.g.\ \SI{10}{\meter \per \second} for $62S$). 

\section{Conclusion}

This work explores the fundamental limit of the lifetime of a Rydberg atom excited in a ultracold, dense gas, which is limited by inelastic collisions of the Rydberg atom with a neutral ground state atom. Two reaction products are observed within the system. Either deeply bound Rb$_2^+$ molecules are created or the Rydberg atoms change their angular momentum. These two reaction product states have been explained from a new theoretical quantum mechanical framework based on the analysis of the reaction pathways including the role of the Rydberg electron. Both channels are equally probable for Rydberg $S$ states at $n=40$. For higher principal quantum numbers, the probability of molecular Rb$_2^+$ production diminishes (\SI{<10}{\percent} for $n\ge90$). The Rydberg electron plays a key role for the interaction, especially because of the $p$-wave scattering resonance leading to the butterfly state in rubidium. It accelerates the inelastic collision compared to the simulated collision time of the ion-neutral system for low principal quantum numbers ($n < 90$) for which the collision time is on the order of a few microseconds at the BEC densities ($10^{14}$\,cm$^{-3}$). For high principal quantum numbers ($n>140$), the inelastic collision time surprisingly exceeds \SI{10}{\us} even in the very dense environment due to the observed threshold behavior for the inelastic collision time between $90 < n < 110$. Hence, Rydberg states at principal quantum numbers above 110 are better suited, e.g., for applications such as quantum optics, for which a high density can benefit the experiment. Similarly, our findings have deep implications for Rydberg-Rydberg interaction-based quantum information processing~\cite{Saffman2010}, in particular by identifying collisions as decoherence sources, which strongly depend on the principal quantum number chosen. An implication of these observed long-lived states for high principal quantum numbers in dense media is the opportunity to image the wave function of a Rydberg atom in-situ \cite{Karpiuk2015}. The findings of this work can be used to understand the decoherence limits of utilizing Rydberg atoms in cold, dense atom samples. Most interesting will be future studies of the collisional lifetimes of Rydberg atoms using atomic species, which do not have a \textit{p}-wave shape resonance for electron-neutral collisions, for example strontium. Based on the collision mechanisms described in this work, the collisional lifetime depends on the initial temperature of the neutral atom reagent. These studies could be extended by measuring the collisional lifetime of Rydberg atoms in thermal clouds with varying densities and temperatures. 

Finally, the ultimate understanding of the lifetime of Rydberg states in high density media will definitively help to understand the complex intimate interplay between few-body physics and many-body physics, which usually seems to be the most applied approach to these systems.

\begin{acknowledgments}
We thank Andreas K{\"o}hn and David Peter for helpful discussions. J.~P.-R.\ and C.~H.~G.\ acknowledge fruitful discussion with Francis Robicheaux and Matt Eieles. This work was supported by the Deutsche Forschungsgemeinschaft (DFG) within the SFB/TRR21 and the project PF 381/13-1. Parts of this work was also founded by ERC under contract number 267100. S.H.\ acknowledges support from DFG through the project HO 4787/1-1, M.S.\ acknowledges support from the Carl Zeiss Foundation. This work was supported by the Department of Energy, Office of Science, under Award No.\ DE-SC0010545.
\end{acknowledgments}


\appendix
\section{Energy release of the \lc{}-changing collision}
\label{a:EnergyRelease}
\begin{figure}
\begin{center}
\begingroup%
  \makeatletter%
  \providecommand\color[2][]{%
    \errmessage{(Inkscape) Color is used for the text in Inkscape, but the package 'color.sty' is not loaded}%
    \renewcommand\color[2][]{}%
  }%
  \providecommand\transparent[1]{%
    \errmessage{(Inkscape) Transparency is used (non-zero) for the text in Inkscape, but the package 'transparent.sty' is not loaded}%
    \renewcommand\transparent[1]{}%
  }%
  \providecommand\rotatebox[2]{#2}%
  \ifx\svgwidth\undefined%
    \setlength{\unitlength}{202.75981835bp}%
    \ifx\svgscale\undefined%
      \relax%
    \else%
      \setlength{\unitlength}{\unitlength * \real{\svgscale}}%
    \fi%
  \else%
    \setlength{\unitlength}{\svgwidth}%
  \fi%
  \global\let\svgwidth\undefined%
  \global\let\svgscale\undefined%
  \makeatother%
  \begin{picture}(1,0.46923913)%
    \put(0,0){\includegraphics[width=\unitlength,page=1]{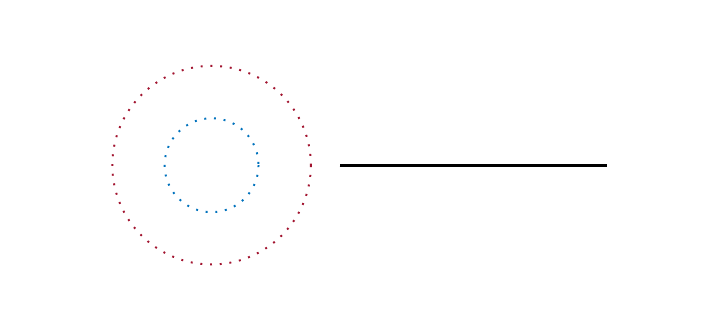}}%
    \put(0.96977349,0.1615649){\color[rgb]{0,0,0}\rotatebox{90}{\makebox(0,0)[lb]{\smash{MCP}}}}%
    \put(0,0){\includegraphics[width=\unitlength,page=2]{./SchematicSpread2.pdf}}%
    \put(0.58788768,0.36271499){\color[rgb]{0,0,0}\makebox(0,0)[lb]{\smash{\shortstack{field\\plate}}}}%
    \put(0,0){\includegraphics[width=\unitlength,page=3]{./SchematicSpread2.pdf}}%
    \put(0,0){\includegraphics[width=\unitlength,page=4]{./SchematicSpread2.pdf}}%
    \put(0.99919676,0.22019241){\color[rgb]{0,0,0}\makebox(0,0)[lb]{\smash{}}}%
    \put(0,0.36271475){\color[rgb]{0,0,0}\makebox(0,0)[lb]{\smash{\shortstack{Rydberg\\atoms}}}}%
  \end{picture}%
\endgroup\\%
\includegraphics{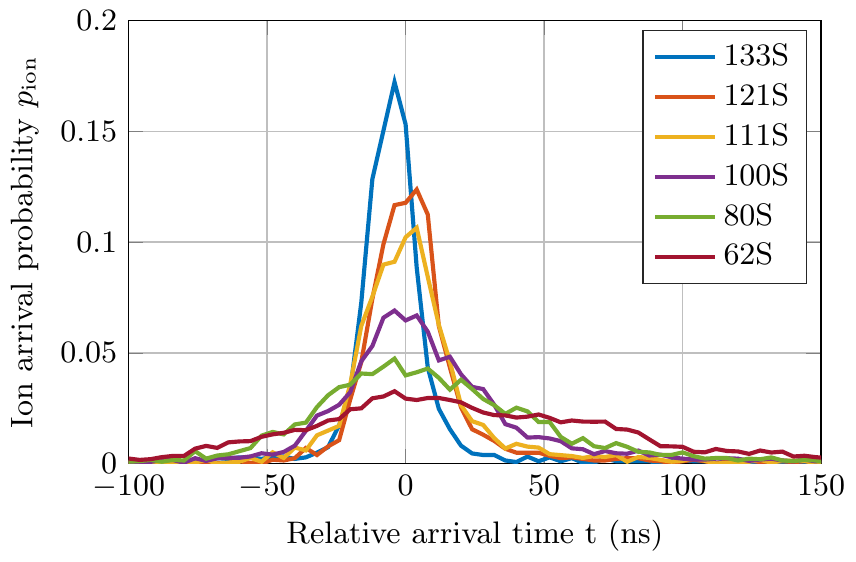}
\caption{\label{fig:IonSpread} Ion arrival time spread due to the \lc-changing collision. The top panel shows schematically the positions of the Rydberg atoms at the moment of the ionization. The Rydberg atoms are located on a sphere with the radius $r(E)$, which depends on the released energy $E$. The bottom panel shows the relative ion arrival time of the ionized Rydberg atoms which went through an \lc{}-changing collision. The highest principal quantum number (dark blue) shows the lowest spread, which means that the energy release during the collision was low. The lowest principal quantum number (dark red) shows the widest spread because of the higher energy release during the inelastic collision. All Rydberg atoms are ionized after a \SI{50}{\us} delay time with an electric field of \SI{180}{\Vcm}.}
\end{center}
\end{figure}The released energy during an \lc{}-changing collision, as shown in Fig.~\sref[a]{fig:nlchange}, can be extracted out of the arrival time spread of the ionized Rydberg atoms on the detector. The method will be explained by the means of Fig.~\ref{fig:IonSpread}, which shows schematically ion flight trajectories leading to the ion arrival time spread. In addition, a measurement series of the ion arrival time spread for different principal quantum number is shown in this figure. All Rydberg atoms were ionized with the same ionization electric field (\SI{180}{\Vcm}) and a constant delay time between the excitation and the ionization (\SI{50}{\us}), so that the spread of the ion arrival time for different principal quantum numbers can be directly compared.
The full width at half maximum of the ion arrival time spread increases from \SI{\approx20}{\us} for the highest state shown, $n=133$, to \SI{\approx100}{\us} for $n=62$. 
The ion arrival time distribution $p_r(t)$ can be simulated as a parameter of the sphere radius $r$, which is the distance the Rydberg atoms have traveled before the ionization with respect to the excitation center $r=0$. Due to the electric field gradients inside the experiment chamber and the asymmetric distance distribution from the ion detector to the Rydberg atoms, the ion arrival time is asymmetric. The ion arrival time distribution of Fig.~\ref{fig:IonSpread} is reconstructed by fitting a linear combination of the simulated arrival time distributions $p_r(t)$ to the measured arrival time distribution $p_\text{ion}(t)$:
\begin{align}
p_\text{ion}(t) = \sum_r c_r p_r(t).
\end{align}
The fit coefficients $c_r$ reflect the probability, $p(r)$, that a Rydberg atom has traveled the distance $r$ in the time between the \lc{}-changing collision and the ionization. This spatial distribution can be converted into a distribution of the released energy $p(E)$ with Eq.~(\ref{eq:EnergyRelease}). The maximum and the full width at half maximum of the energy distribution $p(E)$ are finally plotted in Fig.~\sref[a]{fig:nlchange} and the most probable total energy release corresponds for lower principal quantum numbers, for which the energy release is high, to the energy difference of the initially excited state to the next lower lying manifold. The ion arrival time spread was measured not only for \SI{50}{\us} but also for higher delay times (up to \SI{150}{\us}) to increase the sensitivity especially for high principal quantum numbers, for which the released energy is low.

The method to simulate and extract information out of the ion arrival time was also used recently in another context by Faoro and coworkers to measure in real space the effect of the van der Waals forces between individual Rydberg atoms~\cite{Faoro2016}.

\section{Measuring the \lc{}-changing collision time}
\label{a:Tau}
\begin{figure}
\begin{center}
\includegraphics{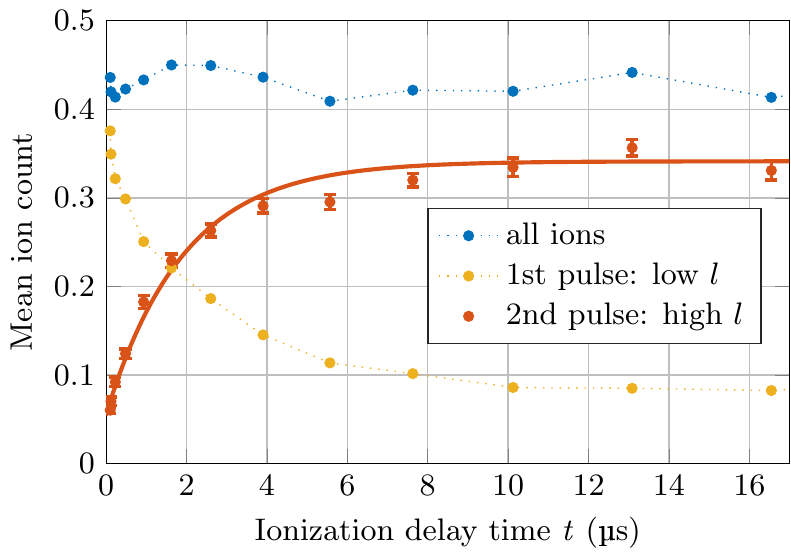}
\caption{\label{fig:TauRb100S} Analysis of the \lc{}-changing collision time for a 100S state at a laser detuning of \SI{-48}{\MHz}. The Rydberg atoms were excited for \SI{200}{\ns} and the first ionization pulse of \SI{10}{\Vcm} mainly ionizes the Rydberg atoms (orange) which are in the initial state. The subsequent higher ionization pulse (\SI{180}{\Vcm}) ionizes the remaining Rydberg atoms which went through the \lc{}-changing collision (red). The sum of both pulses (blue) stays constant within the experimental uncertainty. The mean collision time is extracted out of the red fit curve $P(t) \propto 1 - \exp(-t/\tau)$ as \SI{1.9\pm0.5}{\us} and the error given is coming from the confidence interval of the fit. The error bar of every data point is showing the standard error from the mean from 32 averages.}
\end{center}
\end{figure}
The time of an inelastic collision is determined by measuring the reaction products as a function of the delay time $t$ between the excitation and ionization. An exemplary analysis is shown for the 100S state at a laser detuning of \SI{-48}{\MHz} in Fig.~\ref{fig:TauRb100S}. After a variable delay time $t$, a first ionization pulse (\SI{10}{\Vcm}) is applied to ionize the 100S Rydberg atoms, but not the Rydberg atoms which went through an \lc{}-changing. The second ionization pulse (\SI{180}{\Vcm}) ionizes the remaining Rydberg atoms, which must be as a consequence in a high \lc{} state. The measured signal is not zero at $t=0$ because Rydberg atoms can already collide inelastically during the time of the excitation (\SI{200}{\ns}). The time constant $\tau$ of the collision is fitted with
\begin{align}
P(t) \propto 1-\exp\left(-\frac{t}{\tau}\right)
\end{align}
which results for the examined state in a collision time of $\tau = \SI{1.9\pm0.5}{\us}$. The sum of the ions detected in both ionization pulses is constant within the experimental uncertainty, which shows that all excited Rydberg atoms are detected independent from the delay time before the ionization.

\section{Landau-Zener probabilities}
\label{a:LZ}
\begin{figure}
\begin{center}
\includegraphics{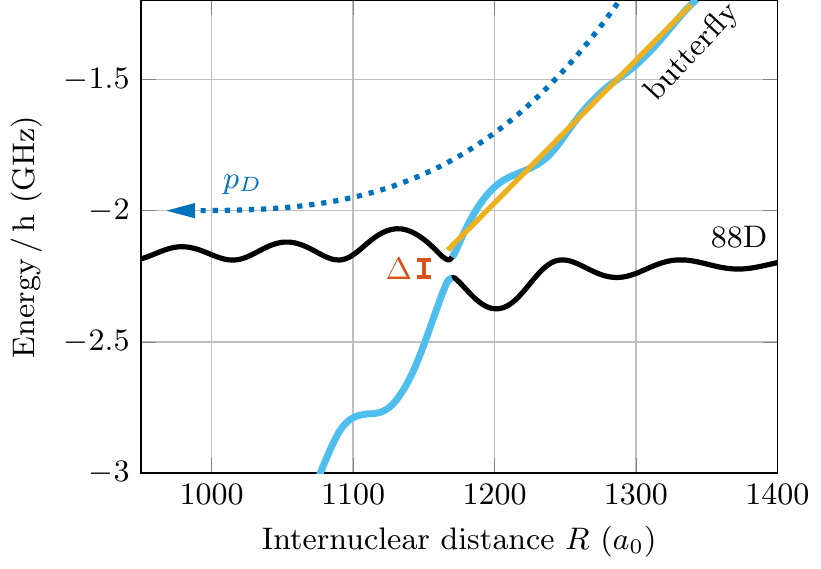}
\caption{\label{Zoom} Avoided crossing between the butterfly state (blue) and the 88$D$ state (black). This avoided crossing is studied by means of a Landau-Zener approach with Eq.~(\ref{eq:LZ}), where $\Delta$ denotes the energy gap at the crossing point and the orange line depicts the slope at the crossing point.}
\end{center}
\end{figure}
In the Landau-Zener framework the non-adiabatic (na) transition probability is given by 
\begin{equation}
\label{eq:LZ}
p_{\text{na}}=\exp\left(\frac{-\pi\Delta^2}{2 \alpha v}\right),
\end{equation}
where $v$ denotes the velocity at the crossing point, $\alpha$ is the difference between slopes of the diabatic potentials associated with the states under consideration~\cite{Clark1979}, and $\Delta$ denotes the energy difference of the two involved PEC at the crossing point (see Fig.~\ref{Zoom}). In Fig.~\ref{Zoom} the physics behind the Landau-Zener approach is illustrated for the crossing between the butterfly states and the 88$D$ state, as it was already shown in Fig.~\ref{fig:PEC}. The probability to end up in the 88$D$ state is given by Eq.~(\ref{eq:LZ}), where the velocity at the crossing point in atomic units is given by 
\begin{equation}
v=\sqrt{\frac{1}{m_{\text{Rb}}}\left(\frac{2}{(90-\delta_{S})^2}-\frac{2}{(88-\delta_{D})^2} \right)}.
\end{equation}
For each Landau-Zener crossing an error bar has been estimated assuming a deviation $\sim$ 20 $\%$ in $\alpha$ and 10 $\%$ in $\Delta$, and the results are shown in panel (b) of Fig.~\ref{fig:PEC} of the main text.

\section{Trilobite-low angular momentum states couplings}
\label{a:pmatrix}
\begin{figure}
\begin{center}
\includegraphics{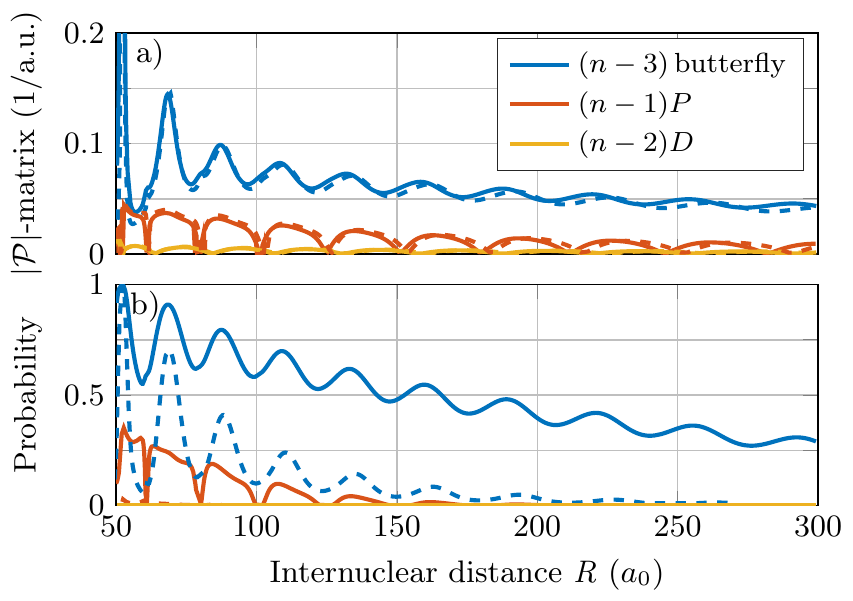}
\caption{(a)~\label{fig:pmatrix}$\mathcal{P}$-matrix elements associated with the butterfly-trilobite coupling (blue), $(n-1)P$-trilobite (red) and $(n-2)D$-trilobite crossing (orange) for two different principal quantum numbers, $n = 40$ (dashed lines) and $n = 90$ (solid lines). (b)~Probability for non-adiabatic transitions associated with the $\mathcal{P}$-matrix displayed in panel (a), see text for details.}
\end{center}
\end{figure}

The coupling between the trilobite state and the rest of the involved Rydberg states has been calculated by means of the nonadiabatic coupling $\mathcal{P}$-matrix defined
\begin{equation}
\mathcal{P}_{ij}(R)=\langle \Phi_i(R)|\frac{\partial}{\partial R}|\Phi_j(R) \rangle,
\end{equation}
where $|\Phi_j(R) \rangle$ and $|\Phi_i(R) \rangle$ represent the adiabatic states involved. The $\mathcal{P}$-matrix associated with the couplings between the trilobite state and the relevant states for $l$-mixing collisions are shown for two different quantum numbers $n$ = 40 and $n$ = 90 in panel (a) of Fig.~\ref{fig:pmatrix}. Observe that the $(n-2)D$ state shows a negligible coupling with respect to the trilobite state, and hence every atom ending up in the $(n-2)D$ state after passing through the butterfly region will reach the short-range region dominated by chemical forces. However, atoms in the $(n-1)P$ state or butterfly state after the butterfly region will experience a considerable coupling with the trilobite state correlated with the $(n-4)$ hydrogenic manifold. 

The probability for non-adiabatic transitions can be calculated in atomic units as \cite{Clark1979}
\begin{equation}
\label{d2}
p_{\text{na}}=\exp\left(-2\pi \xi\right),
\end{equation}
where $\xi=\Delta/(8v\mathcal{P}_{\text{max}})$~\cite{Clark1979}, and $v$ denotes the velocity at the crossing point, $\Delta$ stands for the energy difference between the involved adiabatic states, and $\mathcal{P}_{\text{max}}$ denotes the value of the $\mathcal{P}$-matrix at the crossing point, which corresponds to its maximum value. As it is noticed in panel (a) of Fig.~\ref{fig:pmatrix}, some of the observed crossings (regions where the $\mathcal{P}$-matrix shows a local maximum) are close to each other. Therefore, Eq.~(\ref{d2}) has been extended to all distances, i.e.\ $\xi=\Delta/(8v\mathcal{P})$~\cite{Clark1979}, where $\mathcal{P}$ denotes the $\mathcal{P}$-matrix associated with the cited adiabatic states. In panel (b) of Fig.~\ref{fig:pmatrix}, the non-adiabatic transition probability for the crossings in panel (a) Fig.~\ref{fig:pmatrix} are shown, where it is clearly observed that the trilobite state is strongly coupled to the butterfly state (blue line). A large coupling between the trilobite state and the $P$ state is also observed (orange line). These couplings are used in the theoretical description of the observed $l$-mixing collisions as discussed in the body of this paper. 


\begin{thebibliography}{49}%
\makeatletter
\providecommand \@ifxundefined [1]{%
 \@ifx{#1\undefined}
}%
\providecommand \@ifnum [1]{%
 \ifnum #1\expandafter \@firstoftwo
 \else \expandafter \@secondoftwo
 \fi
}%
\providecommand \@ifx [1]{%
 \ifx #1\expandafter \@firstoftwo
 \else \expandafter \@secondoftwo
 \fi
}%
\providecommand \natexlab [1]{#1}%
\providecommand \enquote  [1]{``#1''}%
\providecommand \bibnamefont  [1]{#1}%
\providecommand \bibfnamefont [1]{#1}%
\providecommand \citenamefont [1]{#1}%
\providecommand \href@noop [0]{\@secondoftwo}%
\providecommand \href [0]{\begingroup \@sanitize@url \@href}%
\providecommand \@href[1]{\@@startlink{#1}\@@href}%
\providecommand \@@href[1]{\endgroup#1\@@endlink}%
\providecommand \@sanitize@url [0]{\catcode `\\12\catcode `\$12\catcode
  `\&12\catcode `\#12\catcode `\^12\catcode `\_12\catcode `\%12\relax}%
\providecommand \@@startlink[1]{}%
\providecommand \@@endlink[0]{}%
\providecommand \url  [0]{\begingroup\@sanitize@url \@url }%
\providecommand \@url [1]{\endgroup\@href {#1}{\urlprefix }}%
\providecommand \urlprefix  [0]{URL }%
\providecommand \Eprint [0]{\href }%
\providecommand \doibase [0]{http://dx.doi.org/}%
\providecommand \selectlanguage [0]{\@gobble}%
\providecommand \bibinfo  [0]{\@secondoftwo}%
\providecommand \bibfield  [0]{\@secondoftwo}%
\providecommand \translation [1]{[#1]}%
\providecommand \BibitemOpen [0]{}%
\providecommand \bibitemStop [0]{}%
\providecommand \bibitemNoStop [0]{.\EOS\space}%
\providecommand \EOS [0]{\spacefactor3000\relax}%
\providecommand \BibitemShut  [1]{\csname bibitem#1\endcsname}%
\let\auto@bib@innerbib\@empty
\bibitem [{\citenamefont {Heidemann}\ \emph {et~al.}(2007)\citenamefont
  {Heidemann}, \citenamefont {Raitzsch}, \citenamefont {Bendkowsky},
  \citenamefont {Butscher}, \citenamefont {L{\"{o}}w}, \citenamefont {Santos},\
  and\ \citenamefont {Pfau}}]{Heidemann2007}%
  \BibitemOpen
  \bibfield  {author} {\bibinfo {author} {\bibfnamefont {R.}~\bibnamefont
  {Heidemann}}, \bibinfo {author} {\bibfnamefont {U.}~\bibnamefont {Raitzsch}},
  \bibinfo {author} {\bibfnamefont {V.}~\bibnamefont {Bendkowsky}}, \bibinfo
  {author} {\bibfnamefont {B.}~\bibnamefont {Butscher}}, \bibinfo {author}
  {\bibfnamefont {R.}~\bibnamefont {L{\"{o}}w}}, \bibinfo {author}
  {\bibfnamefont {L.}~\bibnamefont {Santos}}, \ and\ \bibinfo {author}
  {\bibfnamefont {T.}~\bibnamefont {Pfau}},\ }\bibfield  {title} {\enquote
  {\bibinfo {title} {{Evidence for coherent collective Rydberg excitation in
  the strong blockade regime.}}}\ }\href {\doibase
  10.1103/PhysRevLett.99.163601} {\bibfield  {journal} {\bibinfo  {journal}
  {Phys. Rev. Lett.}\ }\textbf {\bibinfo {volume} {99}},\ \bibinfo {pages}
  {163601} (\bibinfo {year} {2007})}\BibitemShut {NoStop}%
\bibitem [{\citenamefont {Dudin}\ \emph {et~al.}(2012)\citenamefont {Dudin},
  \citenamefont {Li}, \citenamefont {Bariani},\ and\ \citenamefont
  {Kuzmich}}]{Dudin2012}%
  \BibitemOpen
  \bibfield  {author} {\bibinfo {author} {\bibfnamefont {Y.~O.}\ \bibnamefont
  {Dudin}}, \bibinfo {author} {\bibfnamefont {L.}~\bibnamefont {Li}}, \bibinfo
  {author} {\bibfnamefont {F.}~\bibnamefont {Bariani}}, \ and\ \bibinfo
  {author} {\bibfnamefont {A.}~\bibnamefont {Kuzmich}},\ }\bibfield  {title}
  {\enquote {\bibinfo {title} {{Observation of coherent many-body Rabi
  oscillations}},}\ }\href {\doibase 10.1038/nphys2413} {\bibfield  {journal}
  {\bibinfo  {journal} {Nat. Phys.}\ }\textbf {\bibinfo {volume} {8}},\
  \bibinfo {pages} {790} (\bibinfo {year} {2012})}\BibitemShut {NoStop}%
\bibitem [{\citenamefont {Ebert}\ \emph {et~al.}(2014)\citenamefont {Ebert},
  \citenamefont {Gill}, \citenamefont {Gibbons}, \citenamefont {Zhang},
  \citenamefont {Saffman},\ and\ \citenamefont {Walker}}]{Ebert2014}%
  \BibitemOpen
  \bibfield  {author} {\bibinfo {author} {\bibfnamefont {M.}~\bibnamefont
  {Ebert}}, \bibinfo {author} {\bibfnamefont {A.}~\bibnamefont {Gill}},
  \bibinfo {author} {\bibfnamefont {M.}~\bibnamefont {Gibbons}}, \bibinfo
  {author} {\bibfnamefont {X.}~\bibnamefont {Zhang}}, \bibinfo {author}
  {\bibfnamefont {M.}~\bibnamefont {Saffman}}, \ and\ \bibinfo {author}
  {\bibfnamefont {T.~G.}\ \bibnamefont {Walker}},\ }\bibfield  {title}
  {\enquote {\bibinfo {title} {{Atomic Fock state preparation using Rydberg
  blockade.}}}\ }\href {\doibase 10.1103/PhysRevLett.112.043602} {\bibfield
  {journal} {\bibinfo  {journal} {Phys. Rev. Lett.}\ }\textbf {\bibinfo
  {volume} {112}},\ \bibinfo {pages} {043602} (\bibinfo {year}
  {2014})}\BibitemShut {NoStop}%
\bibitem [{\citenamefont {Zeiher}\ \emph {et~al.}(2015)\citenamefont {Zeiher},
  \citenamefont {Schau{\ss}}, \citenamefont {Hild}, \citenamefont
  {Macr{\`{i}}}, \citenamefont {Bloch},\ and\ \citenamefont
  {Gross}}]{Zeiher2015}%
  \BibitemOpen
  \bibfield  {author} {\bibinfo {author} {\bibfnamefont {J.}~\bibnamefont
  {Zeiher}}, \bibinfo {author} {\bibfnamefont {P.}~\bibnamefont {Schau{\ss}}},
  \bibinfo {author} {\bibfnamefont {S.}~\bibnamefont {Hild}}, \bibinfo {author}
  {\bibfnamefont {T.}~\bibnamefont {Macr{\`{i}}}}, \bibinfo {author}
  {\bibfnamefont {I.}~\bibnamefont {Bloch}}, \ and\ \bibinfo {author}
  {\bibfnamefont {C.}~\bibnamefont {Gross}},\ }\bibfield  {title} {\enquote
  {\bibinfo {title} {{Microscopic Characterization of Scalable Coherent Rydberg
  Superatoms}},}\ }\href {\doibase 10.1103/PhysRevX.5.031015} {\bibfield
  {journal} {\bibinfo  {journal} {Phys. Rev. X}\ }\textbf {\bibinfo {volume}
  {5}},\ \bibinfo {pages} {031015} (\bibinfo {year} {2015})}\BibitemShut
  {NoStop}%
\bibitem [{\citenamefont {G{\"{u}}nter}\ \emph {et~al.}(2013)\citenamefont
  {G{\"{u}}nter}, \citenamefont {Schempp}, \citenamefont {Robert-de
  Saint-Vincent}, \citenamefont {Gavryusev}, \citenamefont {Helmrich},
  \citenamefont {Hofmann}, \citenamefont {Whitlock},\ and\ \citenamefont
  {Weidem{\"{u}}ller}}]{Guenter2013}%
  \BibitemOpen
  \bibfield  {author} {\bibinfo {author} {\bibfnamefont {G.}~\bibnamefont
  {G{\"{u}}nter}}, \bibinfo {author} {\bibfnamefont {H.}~\bibnamefont
  {Schempp}}, \bibinfo {author} {\bibfnamefont {M.}~\bibnamefont {Robert-de
  Saint-Vincent}}, \bibinfo {author} {\bibfnamefont {V.}~\bibnamefont
  {Gavryusev}}, \bibinfo {author} {\bibfnamefont {S.}~\bibnamefont {Helmrich}},
  \bibinfo {author} {\bibfnamefont {C.~S.}\ \bibnamefont {Hofmann}}, \bibinfo
  {author} {\bibfnamefont {S.}~\bibnamefont {Whitlock}}, \ and\ \bibinfo
  {author} {\bibfnamefont {M.}~\bibnamefont {Weidem{\"{u}}ller}},\ }\bibfield
  {title} {\enquote {\bibinfo {title} {{Observing the dynamics of
  dipole-mediated energy transport by interaction-enhanced imaging.}}}\ }\href
  {\doibase 10.1126/science.1244843} {\bibfield  {journal} {\bibinfo  {journal}
  {Science}\ }\textbf {\bibinfo {volume} {342}},\ \bibinfo {pages} {954}
  (\bibinfo {year} {2013})}\BibitemShut {NoStop}%
\bibitem [{\citenamefont {Zeiher}\ \emph {et~al.}(2016)\citenamefont {Zeiher},
  \citenamefont {van Bijnen}, \citenamefont {Schau{\ss}}, \citenamefont {Hild},
  \citenamefont {Choi}, \citenamefont {Pohl}, \citenamefont {Bloch},\ and\
  \citenamefont {Gross}}]{Zeiher2016}%
  \BibitemOpen
  \bibfield  {author} {\bibinfo {author} {\bibfnamefont {J.}~\bibnamefont
  {Zeiher}}, \bibinfo {author} {\bibfnamefont {R.}~\bibnamefont {van Bijnen}},
  \bibinfo {author} {\bibfnamefont {P.}~\bibnamefont {Schau{\ss}}}, \bibinfo
  {author} {\bibfnamefont {S.}~\bibnamefont {Hild}}, \bibinfo {author}
  {\bibfnamefont {J.-y.}\ \bibnamefont {Choi}}, \bibinfo {author}
  {\bibfnamefont {T.}~\bibnamefont {Pohl}}, \bibinfo {author} {\bibfnamefont
  {I.}~\bibnamefont {Bloch}}, \ and\ \bibinfo {author} {\bibfnamefont
  {C.}~\bibnamefont {Gross}},\ }\bibfield  {title} {\enquote {\bibinfo {title}
  {{Many-body interferometry of a Rydberg-dressed spin lattice}},}\ }\href
  {http://arxiv.org/abs/1602.06313} {\bibfield  {journal} {\bibinfo  {journal}
  {arXiv}\ } (\bibinfo {year} {2016})},\ \Eprint
  {http://arxiv.org/abs/1602.06313} {arXiv:1602.06313} \BibitemShut {NoStop}%
\bibitem [{\citenamefont {Baur}\ \emph {et~al.}(2014)\citenamefont {Baur},
  \citenamefont {Tiarks}, \citenamefont {Rempe},\ and\ \citenamefont
  {D{\"{u}}rr}}]{Baur2014}%
  \BibitemOpen
  \bibfield  {author} {\bibinfo {author} {\bibfnamefont {S.}~\bibnamefont
  {Baur}}, \bibinfo {author} {\bibfnamefont {D.}~\bibnamefont {Tiarks}},
  \bibinfo {author} {\bibfnamefont {G.}~\bibnamefont {Rempe}}, \ and\ \bibinfo
  {author} {\bibfnamefont {S.}~\bibnamefont {D{\"{u}}rr}},\ }\bibfield  {title}
  {\enquote {\bibinfo {title} {{Single-photon switch based on Rydberg
  blockade.}}}\ }\href {\doibase 10.1103/PhysRevLett.112.073901} {\bibfield
  {journal} {\bibinfo  {journal} {Phys. Rev. Lett.}\ }\textbf {\bibinfo
  {volume} {112}},\ \bibinfo {pages} {073901} (\bibinfo {year}
  {2014})}\BibitemShut {NoStop}%
\bibitem [{\citenamefont {Gorniaczyk}\ \emph {et~al.}(2014)\citenamefont
  {Gorniaczyk}, \citenamefont {Tresp}, \citenamefont {Schmidt}, \citenamefont
  {Fedder},\ and\ \citenamefont {Hofferberth}}]{Gorniaczyk2014}%
  \BibitemOpen
  \bibfield  {author} {\bibinfo {author} {\bibfnamefont {H.}~\bibnamefont
  {Gorniaczyk}}, \bibinfo {author} {\bibfnamefont {C.}~\bibnamefont {Tresp}},
  \bibinfo {author} {\bibfnamefont {J.}~\bibnamefont {Schmidt}}, \bibinfo
  {author} {\bibfnamefont {H.}~\bibnamefont {Fedder}}, \ and\ \bibinfo {author}
  {\bibfnamefont {S.}~\bibnamefont {Hofferberth}},\ }\bibfield  {title}
  {\enquote {\bibinfo {title} {{Single-photon transistor mediated by interstate
  Rydberg interactions}},}\ }\href {\doibase 10.1103/PhysRevLett.113.053601}
  {\bibfield  {journal} {\bibinfo  {journal} {Phys. Rev. Lett.}\ }\textbf
  {\bibinfo {volume} {113}},\ \bibinfo {pages} {053601} (\bibinfo {year}
  {2014})}\BibitemShut {NoStop}%
\bibitem [{\citenamefont {Tiarks}\ \emph {et~al.}(2014)\citenamefont {Tiarks},
  \citenamefont {Baur}, \citenamefont {Schneider}, \citenamefont {D{\"{u}}rr},\
  and\ \citenamefont {Rempe}}]{Tiarks2014}%
  \BibitemOpen
  \bibfield  {author} {\bibinfo {author} {\bibfnamefont {D.}~\bibnamefont
  {Tiarks}}, \bibinfo {author} {\bibfnamefont {S.}~\bibnamefont {Baur}},
  \bibinfo {author} {\bibfnamefont {K.}~\bibnamefont {Schneider}}, \bibinfo
  {author} {\bibfnamefont {S.}~\bibnamefont {D{\"{u}}rr}}, \ and\ \bibinfo
  {author} {\bibfnamefont {G.}~\bibnamefont {Rempe}},\ }\bibfield  {title}
  {\enquote {\bibinfo {title} {{Single-photon transistor using a F{\"{o}}rster
  resonance}},}\ }\href {\doibase 10.1103/PhysRevLett.113.053602} {\bibfield
  {journal} {\bibinfo  {journal} {Phys. Rev. Lett.}\ }\textbf {\bibinfo
  {volume} {113}},\ \bibinfo {pages} {053602} (\bibinfo {year}
  {2014})}\BibitemShut {NoStop}%
\bibitem [{\citenamefont {Greene}\ \emph {et~al.}(2000)\citenamefont {Greene},
  \citenamefont {Dickinson},\ and\ \citenamefont {Sadeghpour}}]{Greene2000}%
  \BibitemOpen
  \bibfield  {author} {\bibinfo {author} {\bibfnamefont {C.~H.}\ \bibnamefont
  {Greene}}, \bibinfo {author} {\bibfnamefont {A.~S.}\ \bibnamefont
  {Dickinson}}, \ and\ \bibinfo {author} {\bibfnamefont {H.~R.}\ \bibnamefont
  {Sadeghpour}},\ }\bibfield  {title} {\enquote {\bibinfo {title} {{Creation of
  Polar and Nonpolar Ultra-Long-Range Rydberg Molecules}},}\ }\href {\doibase
  10.1103/PhysRevLett.85.2458} {\bibfield  {journal} {\bibinfo  {journal}
  {Phys. Rev. Lett.}\ }\textbf {\bibinfo {volume} {85}},\ \bibinfo {pages}
  {2458} (\bibinfo {year} {2000})}\BibitemShut {NoStop}%
\bibitem [{\citenamefont {Bendkowsky}\ \emph {et~al.}(2009)\citenamefont
  {Bendkowsky}, \citenamefont {Butscher}, \citenamefont {Nipper}, \citenamefont
  {Shaffer}, \citenamefont {L{\"{o}}w},\ and\ \citenamefont
  {Pfau}}]{Bendkowsky2009}%
  \BibitemOpen
  \bibfield  {author} {\bibinfo {author} {\bibfnamefont {V.}~\bibnamefont
  {Bendkowsky}}, \bibinfo {author} {\bibfnamefont {B.}~\bibnamefont
  {Butscher}}, \bibinfo {author} {\bibfnamefont {J.}~\bibnamefont {Nipper}},
  \bibinfo {author} {\bibfnamefont {J.~P.}\ \bibnamefont {Shaffer}}, \bibinfo
  {author} {\bibfnamefont {R.}~\bibnamefont {L{\"{o}}w}}, \ and\ \bibinfo
  {author} {\bibfnamefont {T.}~\bibnamefont {Pfau}},\ }\bibfield  {title}
  {\enquote {\bibinfo {title} {{Observation of ultralong-range Rydberg
  molecules}},}\ }\href {\doibase 10.1038/nature07945} {\bibfield  {journal}
  {\bibinfo  {journal} {Nature}\ }\textbf {\bibinfo {volume} {458}},\ \bibinfo
  {pages} {1005} (\bibinfo {year} {2009})}\BibitemShut {NoStop}%
\bibitem [{\citenamefont {Niederpr{\"{u}}m}\ \emph {et~al.}(2016)\citenamefont
  {Niederpr{\"{u}}m}, \citenamefont {Thomas}, \citenamefont {Eichert},
  \citenamefont {Lippe}, \citenamefont {P{\'{e}}rez-R{\'{i}}os}, \citenamefont
  {Greene},\ and\ \citenamefont {Ott}}]{Niederpruem2016}%
  \BibitemOpen
  \bibfield  {author} {\bibinfo {author} {\bibfnamefont {T.}~\bibnamefont
  {Niederpr{\"{u}}m}}, \bibinfo {author} {\bibfnamefont {O.}~\bibnamefont
  {Thomas}}, \bibinfo {author} {\bibfnamefont {T.}~\bibnamefont {Eichert}},
  \bibinfo {author} {\bibfnamefont {C.}~\bibnamefont {Lippe}}, \bibinfo
  {author} {\bibfnamefont {J.}~\bibnamefont {P{\'{e}}rez-R{\'{i}}os}}, \bibinfo
  {author} {\bibfnamefont {C.~H.}\ \bibnamefont {Greene}}, \ and\ \bibinfo
  {author} {\bibfnamefont {H.}~\bibnamefont {Ott}},\ }\bibfield  {title}
  {\enquote {\bibinfo {title} {{Observation of pendular butterfly Rydberg
  molecules}},}\ }\href {http://arxiv.org/abs/1602.08400} {\bibfield  {journal}
  {\bibinfo  {journal} {arXiv}\ } (\bibinfo {year} {2016})},\ \Eprint
  {http://arxiv.org/abs/1602.08400} {arXiv:1602.08400} \BibitemShut {NoStop}%
\bibitem [{\citenamefont {Li}\ \emph {et~al.}(2011)\citenamefont {Li},
  \citenamefont {Pohl}, \citenamefont {Rost}, \citenamefont {Rittenhouse},
  \citenamefont {Sadeghpour}, \citenamefont {Nipper}, \citenamefont {Butscher},
  \citenamefont {Balewski}, \citenamefont {Bendkowsky}, \citenamefont
  {L{\"{o}}w},\ and\ \citenamefont {Pfau}}]{Li2011}%
  \BibitemOpen
  \bibfield  {author} {\bibinfo {author} {\bibfnamefont {W.}~\bibnamefont
  {Li}}, \bibinfo {author} {\bibfnamefont {T.}~\bibnamefont {Pohl}}, \bibinfo
  {author} {\bibfnamefont {J.~M.}\ \bibnamefont {Rost}}, \bibinfo {author}
  {\bibfnamefont {S.~T.}\ \bibnamefont {Rittenhouse}}, \bibinfo {author}
  {\bibfnamefont {H.~R.}\ \bibnamefont {Sadeghpour}}, \bibinfo {author}
  {\bibfnamefont {J.}~\bibnamefont {Nipper}}, \bibinfo {author} {\bibfnamefont
  {B.}~\bibnamefont {Butscher}}, \bibinfo {author} {\bibfnamefont {J.~B.}\
  \bibnamefont {Balewski}}, \bibinfo {author} {\bibfnamefont {V.}~\bibnamefont
  {Bendkowsky}}, \bibinfo {author} {\bibfnamefont {R.}~\bibnamefont
  {L{\"{o}}w}}, \ and\ \bibinfo {author} {\bibfnamefont {T.}~\bibnamefont
  {Pfau}},\ }\bibfield  {title} {\enquote {\bibinfo {title} {{A homonuclear
  molecule with a permanent electric dipole moment}},}\ }\href {\doibase
  10.1126/science.1211255} {\bibfield  {journal} {\bibinfo  {journal}
  {Science}\ }\textbf {\bibinfo {volume} {334}},\ \bibinfo {pages} {1110}
  (\bibinfo {year} {2011})}\BibitemShut {NoStop}%
\bibitem [{\citenamefont {Balewski}\ \emph {et~al.}(2013)\citenamefont
  {Balewski}, \citenamefont {Krupp}, \citenamefont {Gaj}, \citenamefont
  {Peter}, \citenamefont {B{\"{u}}chler}, \citenamefont {L{\"{o}}w},
  \citenamefont {Hofferberth},\ and\ \citenamefont {Pfau}}]{Balewski2013}%
  \BibitemOpen
  \bibfield  {author} {\bibinfo {author} {\bibfnamefont {J.~B.}\ \bibnamefont
  {Balewski}}, \bibinfo {author} {\bibfnamefont {A.~T.}\ \bibnamefont {Krupp}},
  \bibinfo {author} {\bibfnamefont {A.}~\bibnamefont {Gaj}}, \bibinfo {author}
  {\bibfnamefont {D.}~\bibnamefont {Peter}}, \bibinfo {author} {\bibfnamefont
  {H.~P.}\ \bibnamefont {B{\"{u}}chler}}, \bibinfo {author} {\bibfnamefont
  {R.}~\bibnamefont {L{\"{o}}w}}, \bibinfo {author} {\bibfnamefont
  {S.}~\bibnamefont {Hofferberth}}, \ and\ \bibinfo {author} {\bibfnamefont
  {T.}~\bibnamefont {Pfau}},\ }\bibfield  {title} {\enquote {\bibinfo {title}
  {{Coupling a single electron to a Bose-Einstein condensate}},}\ }\href
  {\doibase 10.1038/nature12592} {\bibfield  {journal} {\bibinfo  {journal}
  {Nature}\ }\textbf {\bibinfo {volume} {502}},\ \bibinfo {pages} {664}
  (\bibinfo {year} {2013})}\BibitemShut {NoStop}%
\bibitem [{\citenamefont {Karpiuk}\ \emph {et~al.}(2015)\citenamefont
  {Karpiuk}, \citenamefont {Brewczyk}, \citenamefont
  {Rz\textpolhook{a}{\.{z}}ewski}, \citenamefont {Gaj}, \citenamefont
  {Balewski}, \citenamefont {Krupp}, \citenamefont {Schlagm{\"{u}}ller},
  \citenamefont {L{\"{o}}w}, \citenamefont {Hofferberth},\ and\ \citenamefont
  {Pfau}}]{Karpiuk2015}%
  \BibitemOpen
  \bibfield  {author} {\bibinfo {author} {\bibfnamefont {T.}~\bibnamefont
  {Karpiuk}}, \bibinfo {author} {\bibfnamefont {M.}~\bibnamefont {Brewczyk}},
  \bibinfo {author} {\bibfnamefont {K.}~\bibnamefont
  {Rz\textpolhook{a}{\.{z}}ewski}}, \bibinfo {author} {\bibfnamefont
  {A.}~\bibnamefont {Gaj}}, \bibinfo {author} {\bibfnamefont {J.~B.}\
  \bibnamefont {Balewski}}, \bibinfo {author} {\bibfnamefont {A.~T.}\
  \bibnamefont {Krupp}}, \bibinfo {author} {\bibfnamefont {M.}~\bibnamefont
  {Schlagm{\"{u}}ller}}, \bibinfo {author} {\bibfnamefont {R.}~\bibnamefont
  {L{\"{o}}w}}, \bibinfo {author} {\bibfnamefont {S.}~\bibnamefont
  {Hofferberth}}, \ and\ \bibinfo {author} {\bibfnamefont {T.}~\bibnamefont
  {Pfau}},\ }\bibfield  {title} {\enquote {\bibinfo {title} {{Imaging single
  Rydberg electrons in a Bose–Einstein condensate}},}\ }\href {\doibase
  10.1088/1367-2630/17/5/053046} {\bibfield  {journal} {\bibinfo  {journal}
  {New J. Phys.}\ }\textbf {\bibinfo {volume} {17}},\ \bibinfo {pages} {53046}
  (\bibinfo {year} {2015})}\BibitemShut {NoStop}%
\bibitem [{\citenamefont {Barbier}\ and\ \citenamefont
  {Cheret}(1987)}]{Barbier1987}%
  \BibitemOpen
  \bibfield  {author} {\bibinfo {author} {\bibfnamefont {L.}~\bibnamefont
  {Barbier}}\ and\ \bibinfo {author} {\bibfnamefont {M.}~\bibnamefont
  {Cheret}},\ }\bibfield  {title} {\enquote {\bibinfo {title} {{Experimental
  study of Penning and Hornbeck-Molnar ionisation of rubidium atoms excited in
  a high s or d level (5d$\ge$nl$\ge$11s)}},}\ }\href {\doibase
  10.1088/0022-3700/20/6/011} {\bibfield  {journal} {\bibinfo  {journal} {J.
  Phys. B At. Mol. Phys.}\ }\textbf {\bibinfo {volume} {20}},\ \bibinfo {pages}
  {1229} (\bibinfo {year} {1987})}\BibitemShut {NoStop}%
\bibitem [{\citenamefont {Beigman}\ and\ \citenamefont
  {Lebedev}(1995)}]{Beigman1995}%
  \BibitemOpen
  \bibfield  {author} {\bibinfo {author} {\bibfnamefont {I.}~\bibnamefont
  {Beigman}}\ and\ \bibinfo {author} {\bibfnamefont {V.}~\bibnamefont
  {Lebedev}},\ }\bibfield  {title} {\enquote {\bibinfo {title} {{Collision
  theory of Rydberg atoms with neutral and charged particles}},}\ }\href
  {\doibase 10.1016/0370-1573(95)00074-Q} {\bibfield  {journal} {\bibinfo
  {journal} {Phys. Rep.}\ }\textbf {\bibinfo {volume} {250}},\ \bibinfo {pages}
  {95} (\bibinfo {year} {1995})}\BibitemShut {NoStop}%
\bibitem [{\citenamefont {Kumar}\ \emph {et~al.}(1999)\citenamefont {Kumar},
  \citenamefont {Saha}, \citenamefont {Weatherford},\ and\ \citenamefont
  {Verma}}]{Kumar1999}%
  \BibitemOpen
  \bibfield  {author} {\bibinfo {author} {\bibfnamefont {A.}~\bibnamefont
  {Kumar}}, \bibinfo {author} {\bibfnamefont {B.}~\bibnamefont {Saha}},
  \bibinfo {author} {\bibfnamefont {C.}~\bibnamefont {Weatherford}}, \ and\
  \bibinfo {author} {\bibfnamefont {S.}~\bibnamefont {Verma}},\ }\bibfield
  {title} {\enquote {\bibinfo {title} {{A systematic study of Hornbeck Molnar
  ionization involving Rydberg alkali atoms}},}\ }\href {\doibase
  10.1016/S0166-1280(99)00134-7} {\bibfield  {journal} {\bibinfo  {journal} {J.
  Mol. Struct. Theochem}\ }\textbf {\bibinfo {volume} {487}},\ \bibinfo {pages}
  {1} (\bibinfo {year} {1999})}\BibitemShut {NoStop}%
\bibitem [{\citenamefont {Mihajlov}\ \emph {et~al.}(2012)\citenamefont
  {Mihajlov}, \citenamefont {Sre{\'{c}}kovi{\'{c}}}, \citenamefont
  {Ignjatovi{\'{c}}},\ and\ \citenamefont {Klyucharev}}]{Mihajlov2012}%
  \BibitemOpen
  \bibfield  {author} {\bibinfo {author} {\bibfnamefont {A.~A.}\ \bibnamefont
  {Mihajlov}}, \bibinfo {author} {\bibfnamefont {V.~A.}\ \bibnamefont
  {Sre{\'{c}}kovi{\'{c}}}}, \bibinfo {author} {\bibfnamefont {L.~M.}\
  \bibnamefont {Ignjatovi{\'{c}}}}, \ and\ \bibinfo {author} {\bibfnamefont
  {A.~N.}\ \bibnamefont {Klyucharev}},\ }\bibfield  {title} {\enquote {\bibinfo
  {title} {{The Chemi-Ionization Processes in Slow Collisions of Rydberg Atoms
  with Ground State Atoms: Mechanism and Applications}},}\ }\href {\doibase
  10.1007/s10876-011-0438-7} {\bibfield  {journal} {\bibinfo  {journal} {J.
  Clust. Sci.}\ }\textbf {\bibinfo {volume} {23}},\ \bibinfo {pages} {47}
  (\bibinfo {year} {2012})}\BibitemShut {NoStop}%
\bibitem [{\citenamefont {Niederpr{\"{u}}m}\ \emph {et~al.}(2015)\citenamefont
  {Niederpr{\"{u}}m}, \citenamefont {Thomas}, \citenamefont {Manthey},
  \citenamefont {Weber},\ and\ \citenamefont {Ott}}]{Niederpruem2015}%
  \BibitemOpen
  \bibfield  {author} {\bibinfo {author} {\bibfnamefont {T.}~\bibnamefont
  {Niederpr{\"{u}}m}}, \bibinfo {author} {\bibfnamefont {O.}~\bibnamefont
  {Thomas}}, \bibinfo {author} {\bibfnamefont {T.}~\bibnamefont {Manthey}},
  \bibinfo {author} {\bibfnamefont {T.~M.}\ \bibnamefont {Weber}}, \ and\
  \bibinfo {author} {\bibfnamefont {H.}~\bibnamefont {Ott}},\ }\bibfield
  {title} {\enquote {\bibinfo {title} {{Giant Cross Section for Molecular Ion
  Formation in Ultracold Rydberg Gases.}}}\ }\href {\doibase
  10.1103/PhysRevLett.115.013003} {\bibfield  {journal} {\bibinfo  {journal}
  {Phys. Rev. Lett.}\ }\textbf {\bibinfo {volume} {115}},\ \bibinfo {pages}
  {013003} (\bibinfo {year} {2015})}\BibitemShut {NoStop}%
\bibitem [{\citenamefont {Miller}(1970)}]{Miller1970}%
  \BibitemOpen
  \bibfield  {author} {\bibinfo {author} {\bibfnamefont {W.~H.}\ \bibnamefont
  {Miller}},\ }\bibfield  {title} {\enquote {\bibinfo {title} {{Theory of
  Penning Ionization. I. Atoms}},}\ }\href {\doibase 10.1063/1.1673523}
  {\bibfield  {journal} {\bibinfo  {journal} {J. Chem. Phys.}\ }\textbf
  {\bibinfo {volume} {52}},\ \bibinfo {pages} {3563} (\bibinfo {year}
  {1970})}\BibitemShut {NoStop}%
\bibitem [{\citenamefont {Gallagher}\ \emph {et~al.}(1975)\citenamefont
  {Gallagher}, \citenamefont {Edelstein},\ and\ \citenamefont
  {Hill}}]{Gallagher1975}%
  \BibitemOpen
  \bibfield  {author} {\bibinfo {author} {\bibfnamefont {T.~F.}\ \bibnamefont
  {Gallagher}}, \bibinfo {author} {\bibfnamefont {S.~A.}\ \bibnamefont
  {Edelstein}}, \ and\ \bibinfo {author} {\bibfnamefont {R.~M.}\ \bibnamefont
  {Hill}},\ }\bibfield  {title} {\enquote {\bibinfo {title} {{Collisional
  Angular Momentum Mixing in Rydberg States of Sodium}},}\ }\href {\doibase
  10.1103/PhysRevLett.35.644} {\bibfield  {journal} {\bibinfo  {journal} {Phys.
  Rev. Lett.}\ }\textbf {\bibinfo {volume} {35}},\ \bibinfo {pages} {644}
  (\bibinfo {year} {1975})}\BibitemShut {NoStop}%
\bibitem [{\citenamefont {Gallagher}\ \emph {et~al.}(1977)\citenamefont
  {Gallagher}, \citenamefont {Edelstein},\ and\ \citenamefont
  {Hill}}]{Gallagher1977}%
  \BibitemOpen
  \bibfield  {author} {\bibinfo {author} {\bibfnamefont {T.~F.}\ \bibnamefont
  {Gallagher}}, \bibinfo {author} {\bibfnamefont {S.~A.}\ \bibnamefont
  {Edelstein}}, \ and\ \bibinfo {author} {\bibfnamefont {R.~M.}\ \bibnamefont
  {Hill}},\ }\bibfield  {title} {\enquote {\bibinfo {title} {{Collisional
  angular-momentum mixing of Rydberg states of Na by He, Ne, and Ar}},}\ }\href
  {\doibase 10.1103/PhysRevA.15.1945} {\bibfield  {journal} {\bibinfo
  {journal} {Phys. Rev. A}\ }\textbf {\bibinfo {volume} {15}},\ \bibinfo
  {pages} {1945} (\bibinfo {year} {1977})}\BibitemShut {NoStop}%
\bibitem [{\citenamefont {Gallagher}\ \emph {et~al.}(1978)\citenamefont
  {Gallagher}, \citenamefont {Cooke},\ and\ \citenamefont
  {Edelstein}}]{Gallagher1978}%
  \BibitemOpen
  \bibfield  {author} {\bibinfo {author} {\bibfnamefont {T.~F.}\ \bibnamefont
  {Gallagher}}, \bibinfo {author} {\bibfnamefont {W.~E.}\ \bibnamefont
  {Cooke}}, \ and\ \bibinfo {author} {\bibfnamefont {S.~A.}\ \bibnamefont
  {Edelstein}},\ }\bibfield  {title} {\enquote {\bibinfo {title} {{Collisional
  angular momentum mixing of f states of Na}},}\ }\href {\doibase
  10.1103/PhysRevA.17.904} {\bibfield  {journal} {\bibinfo  {journal} {Phys.
  Rev. A}\ }\textbf {\bibinfo {volume} {17}},\ \bibinfo {pages} {904} (\bibinfo
  {year} {1978})}\BibitemShut {NoStop}%
\bibitem [{\citenamefont {Hugon}\ \emph {et~al.}(1982)\citenamefont {Hugon},
  \citenamefont {Sayer}, \citenamefont {Fournier},\ and\ \citenamefont
  {Gounand}}]{Hugon1982}%
  \BibitemOpen
  \bibfield  {author} {\bibinfo {author} {\bibfnamefont {M.}~\bibnamefont
  {Hugon}}, \bibinfo {author} {\bibfnamefont {B.}~\bibnamefont {Sayer}},
  \bibinfo {author} {\bibfnamefont {P.~R.}\ \bibnamefont {Fournier}}, \ and\
  \bibinfo {author} {\bibfnamefont {F.}~\bibnamefont {Gounand}},\ }\bibfield
  {title} {\enquote {\bibinfo {title} {{Collisional depopulation of rubidium
  Rydberg levels by rare gases}},}\ }\href {\doibase
  10.1088/0022-3700/15/15/016} {\bibfield  {journal} {\bibinfo  {journal} {J.
  Phys. B At. Mol. Phys.}\ }\textbf {\bibinfo {volume} {15}},\ \bibinfo {pages}
  {2391} (\bibinfo {year} {1982})}\BibitemShut {NoStop}%
\bibitem [{\citenamefont {Gersten}(1976)}]{Gersten1976}%
  \BibitemOpen
  \bibfield  {author} {\bibinfo {author} {\bibfnamefont {J.~I.}\ \bibnamefont
  {Gersten}},\ }\bibfield  {title} {\enquote {\bibinfo {title} {{Theory of
  collisional angular-momentum mixing of Rydberg states}},}\ }\href {\doibase
  10.1103/PhysRevA.14.1354} {\bibfield  {journal} {\bibinfo  {journal} {Phys.
  Rev. A}\ }\textbf {\bibinfo {volume} {14}},\ \bibinfo {pages} {1354}
  (\bibinfo {year} {1976})}\BibitemShut {NoStop}%
\bibitem [{\citenamefont {Olson}(1977)}]{Olson1977}%
  \BibitemOpen
  \bibfield  {author} {\bibinfo {author} {\bibfnamefont {R.~E.}\ \bibnamefont
  {Olson}},\ }\bibfield  {title} {\enquote {\bibinfo {title} {{Theoretical
  excitation transfer cross sections for Rydberg Na ($n^2D \to n^2F$)
  transitions from collision with He, Ne, and Ar}},}\ }\href {\doibase
  10.1103/PhysRevA.15.631} {\bibfield  {journal} {\bibinfo  {journal} {Phys.
  Rev. A}\ }\textbf {\bibinfo {volume} {15}},\ \bibinfo {pages} {631} (\bibinfo
  {year} {1977})}\BibitemShut {NoStop}%
\bibitem [{\citenamefont {Hickman}(1978)}]{Hickman1978}%
  \BibitemOpen
  \bibfield  {author} {\bibinfo {author} {\bibfnamefont {A.~P.}\ \bibnamefont
  {Hickman}},\ }\bibfield  {title} {\enquote {\bibinfo {title} {{Theory of
  angular momentum mixing in Rydberg-atom-rare-gas collisions}},}\ }\href
  {\doibase 10.1103/PhysRevA.18.1339} {\bibfield  {journal} {\bibinfo
  {journal} {Phys. Rev. A}\ }\textbf {\bibinfo {volume} {18}},\ \bibinfo
  {pages} {1339} (\bibinfo {year} {1978})}\BibitemShut {NoStop}%
\bibitem [{\citenamefont {Liebisch}\ \emph {et~al.}(2016)\citenamefont
  {Liebisch}, \citenamefont {Schlagm{\"{u}}ller}, \citenamefont {Engel},
  \citenamefont {Nguyen}, \citenamefont {Balewski}, \citenamefont {Lochead},
  \citenamefont {B{\"{o}}ttcher}, \citenamefont {Westphal}, \citenamefont
  {Kleinbach}, \citenamefont {Gaj}, \citenamefont {L{\"{o}}w}, \citenamefont
  {Hofferberth}, \citenamefont {Pfau}, \citenamefont {P{\'{e}}rez-R\'ios},\
  and\ \citenamefont {Greene}}]{Liebisch2016}%
  \BibitemOpen
  \bibfield  {author} {\bibinfo {author} {\bibfnamefont {T.~C.}\ \bibnamefont
  {Liebisch}}, \bibinfo {author} {\bibfnamefont {M.}~\bibnamefont
  {Schlagm{\"{u}}ller}}, \bibinfo {author} {\bibfnamefont {F.}~\bibnamefont
  {Engel}}, \bibinfo {author} {\bibfnamefont {H.}~\bibnamefont {Nguyen}},
  \bibinfo {author} {\bibfnamefont {J.}~\bibnamefont {Balewski}}, \bibinfo
  {author} {\bibfnamefont {G.}~\bibnamefont {Lochead}}, \bibinfo {author}
  {\bibfnamefont {F.}~\bibnamefont {B{\"{o}}ttcher}}, \bibinfo {author}
  {\bibfnamefont {K.~M.}\ \bibnamefont {Westphal}}, \bibinfo {author}
  {\bibfnamefont {K.~S.}\ \bibnamefont {Kleinbach}}, \bibinfo {author}
  {\bibfnamefont {A.}~\bibnamefont {Gaj}}, \bibinfo {author} {\bibfnamefont
  {R.}~\bibnamefont {L{\"{o}}w}}, \bibinfo {author} {\bibfnamefont
  {S.}~\bibnamefont {Hofferberth}}, \bibinfo {author} {\bibfnamefont
  {T.}~\bibnamefont {Pfau}}, \bibinfo {author} {\bibfnamefont {J.}~\bibnamefont
  {P{\'{e}}rez-R\'ios}}, \ and\ \bibinfo {author} {\bibfnamefont {C.~H.}\
  \bibnamefont {Greene}},\ }\bibfield  {title} {\enquote {\bibinfo {title}
  {{Controlling Rydberg atom excitations in dense background gases}},}\
  }\href@noop {} {\bibfield  {journal} {\bibinfo  {journal} {J. Phys. B, under
  Rev.}\ } (\bibinfo {year} {2016})}\BibitemShut {NoStop}%
\bibitem [{\citenamefont {Esslinger}\ \emph {et~al.}(1998)\citenamefont
  {Esslinger}, \citenamefont {Bloch},\ and\ \citenamefont
  {H{\"{a}}nsch}}]{Esslinger1998}%
  \BibitemOpen
  \bibfield  {author} {\bibinfo {author} {\bibfnamefont {T.}~\bibnamefont
  {Esslinger}}, \bibinfo {author} {\bibfnamefont {I.}~\bibnamefont {Bloch}}, \
  and\ \bibinfo {author} {\bibfnamefont {T.~W.}\ \bibnamefont {H{\"{a}}nsch}},\
  }\bibfield  {title} {\enquote {\bibinfo {title} {{Bose-Einstein condensation
  in a quadrupole-Ioffe-configuration trap}},}\ }\href {\doibase
  10.1103/PhysRevA.58.R2664} {\bibfield  {journal} {\bibinfo  {journal} {Phys.
  Rev. A}\ }\textbf {\bibinfo {volume} {58}},\ \bibinfo {pages} {R2664}
  (\bibinfo {year} {1998})}\BibitemShut {NoStop}%
\bibitem [{\citenamefont {Pethick}\ and\ \citenamefont
  {Smith}(2002)}]{Pethick2002}%
  \BibitemOpen
  \bibfield  {author} {\bibinfo {author} {\bibfnamefont {C.~J.}\ \bibnamefont
  {Pethick}}\ and\ \bibinfo {author} {\bibfnamefont {H.}~\bibnamefont
  {Smith}},\ }\href
  {https://books.google.de/books/about/Bose_Einstein_Condensation_in_Dilute_Gas.html?id=iBk0G3_5iIQC&pgis=1}
  {\emph {\bibinfo {title} {{Bose-Einstein Condensation in Dilute Gases}}}},\
  \bibinfo {edition} {2nd}\ ed.\ (\bibinfo  {publisher} {Cambridge University
  Press},\ \bibinfo {year} {2002})\BibitemShut {NoStop}%
\bibitem [{\citenamefont {Schlagm{\"{u}}ller}\ \emph
  {et~al.}(2016)\citenamefont {Schlagm{\"{u}}ller}, \citenamefont {Liebisch},
  \citenamefont {Nguyen}, \citenamefont {Lochead}, \citenamefont {Engel},
  \citenamefont {B{\"{o}}ttcher}, \citenamefont {Westphal}, \citenamefont
  {Kleinbach}, \citenamefont {L{\"{o}}w}, \citenamefont {Hofferberth},
  \citenamefont {Pfau}, \citenamefont {P{\'{e}}rez-R\'ios},\ and\ \citenamefont
  {Greene}}]{Schlagmueller2016}%
  \BibitemOpen
  \bibfield  {author} {\bibinfo {author} {\bibfnamefont {M.}~\bibnamefont
  {Schlagm{\"{u}}ller}}, \bibinfo {author} {\bibfnamefont {T.~C.}\ \bibnamefont
  {Liebisch}}, \bibinfo {author} {\bibfnamefont {H.}~\bibnamefont {Nguyen}},
  \bibinfo {author} {\bibfnamefont {G.}~\bibnamefont {Lochead}}, \bibinfo
  {author} {\bibfnamefont {F.}~\bibnamefont {Engel}}, \bibinfo {author}
  {\bibfnamefont {F.}~\bibnamefont {B{\"{o}}ttcher}}, \bibinfo {author}
  {\bibfnamefont {K.~M.}\ \bibnamefont {Westphal}}, \bibinfo {author}
  {\bibfnamefont {K.~S.}\ \bibnamefont {Kleinbach}}, \bibinfo {author}
  {\bibfnamefont {R.}~\bibnamefont {L{\"{o}}w}}, \bibinfo {author}
  {\bibfnamefont {S.}~\bibnamefont {Hofferberth}}, \bibinfo {author}
  {\bibfnamefont {T.}~\bibnamefont {Pfau}}, \bibinfo {author} {\bibfnamefont
  {J.}~\bibnamefont {P{\'{e}}rez-R\'ios}}, \ and\ \bibinfo {author}
  {\bibfnamefont {C.~H.}\ \bibnamefont {Greene}},\ }\bibfield  {title}
  {\enquote {\bibinfo {title} {{Probing an Electron Scattering Resonance using
  Rydberg Molecules within a Dense and Ultracold Gas}},}\ }\href {\doibase
  10.1103/PhysRevLett.116.053001} {\bibfield  {journal} {\bibinfo  {journal}
  {Phys. Rev. Lett.}\ }\textbf {\bibinfo {volume} {116}},\ \bibinfo {pages}
  {053001} (\bibinfo {year} {2016})}\BibitemShut {NoStop}%
\bibitem [{\citenamefont {B{\"{o}}ttcher}\ \emph {et~al.}(2016)\citenamefont
  {B{\"{o}}ttcher}, \citenamefont {Gaj}, \citenamefont {Westphal},
  \citenamefont {Schlagm{\"{u}}ller}, \citenamefont {Kleinbach}, \citenamefont
  {L{\"{o}}w}, \citenamefont {Liebisch}, \citenamefont {Pfau},\ and\
  \citenamefont {Hofferberth}}]{Boettcher2016}%
  \BibitemOpen
  \bibfield  {author} {\bibinfo {author} {\bibfnamefont {F.}~\bibnamefont
  {B{\"{o}}ttcher}}, \bibinfo {author} {\bibfnamefont {A.}~\bibnamefont {Gaj}},
  \bibinfo {author} {\bibfnamefont {K.~M.}\ \bibnamefont {Westphal}}, \bibinfo
  {author} {\bibfnamefont {M.}~\bibnamefont {Schlagm{\"{u}}ller}}, \bibinfo
  {author} {\bibfnamefont {K.~S.}\ \bibnamefont {Kleinbach}}, \bibinfo {author}
  {\bibfnamefont {R.}~\bibnamefont {L{\"{o}}w}}, \bibinfo {author}
  {\bibfnamefont {T.~C.}\ \bibnamefont {Liebisch}}, \bibinfo {author}
  {\bibfnamefont {T.}~\bibnamefont {Pfau}}, \ and\ \bibinfo {author}
  {\bibfnamefont {S.}~\bibnamefont {Hofferberth}},\ }\bibfield  {title}
  {\enquote {\bibinfo {title} {{Observation of mixed singlet-triplet Rb$_2$
  Rydberg molecules}},}\ }\href {\doibase 10.1103/PhysRevA.93.032512}
  {\bibfield  {journal} {\bibinfo  {journal} {Phys. Rev. A}\ }\textbf {\bibinfo
  {volume} {93}},\ \bibinfo {pages} {032512} (\bibinfo {year}
  {2016})}\BibitemShut {NoStop}%
\bibitem [{\citenamefont {Gallagher}(1994)}]{Gallagher1994}%
  \BibitemOpen
  \bibfield  {author} {\bibinfo {author} {\bibfnamefont {T.~F.}\ \bibnamefont
  {Gallagher}},\ }\href {\doibase 10.1017/CBO9780511524530} {\emph {\bibinfo
  {title} {{Rydberg Atoms}}}}\ (\bibinfo  {publisher} {Cambridge University
  Press},\ \bibinfo {address} {Cambridge},\ \bibinfo {year} {1994})\BibitemShut
  {NoStop}%
\bibitem [{\citenamefont {Mack}\ \emph {et~al.}(2011)\citenamefont {Mack},
  \citenamefont {Karlewski}, \citenamefont {Hattermann}, \citenamefont
  {H{\"{o}}ckh}, \citenamefont {Jessen}, \citenamefont {Cano},\ and\
  \citenamefont {Fort{\'{a}}gh}}]{Mack2011}%
  \BibitemOpen
  \bibfield  {author} {\bibinfo {author} {\bibfnamefont {M.}~\bibnamefont
  {Mack}}, \bibinfo {author} {\bibfnamefont {F.}~\bibnamefont {Karlewski}},
  \bibinfo {author} {\bibfnamefont {H.}~\bibnamefont {Hattermann}}, \bibinfo
  {author} {\bibfnamefont {S.}~\bibnamefont {H{\"{o}}ckh}}, \bibinfo {author}
  {\bibfnamefont {F.}~\bibnamefont {Jessen}}, \bibinfo {author} {\bibfnamefont
  {D.}~\bibnamefont {Cano}}, \ and\ \bibinfo {author} {\bibfnamefont
  {J.}~\bibnamefont {Fort{\'{a}}gh}},\ }\bibfield  {title} {\enquote {\bibinfo
  {title} {{Measurement of absolute transition frequencies of
  \textsuperscript{87}Rb to \textit{nS} and \textit{nD} Rydberg states by means
  of electromagnetically induced transparency}},}\ }\href {\doibase
  10.1103/PhysRevA.83.052515} {\bibfield  {journal} {\bibinfo  {journal} {Phys.
  Rev. A}\ }\textbf {\bibinfo {volume} {83}},\ \bibinfo {pages} {052515}
  (\bibinfo {year} {2011})}\BibitemShut {NoStop}%
\bibitem [{\citenamefont {Ritz}(1908)}]{Ritz1908}%
  \BibitemOpen
  \bibfield  {author} {\bibinfo {author} {\bibfnamefont {W.}~\bibnamefont
  {Ritz}},\ }\bibfield  {title} {\enquote {\bibinfo {title} {{Magnetische
  Atomfelder und Serienspektren}},}\ }\href {\doibase 10.1002/andp.19083300403}
  {\bibfield  {journal} {\bibinfo  {journal} {Ann. Phys.}\ }\textbf {\bibinfo
  {volume} {330}},\ \bibinfo {pages} {660} (\bibinfo {year}
  {1908})}\BibitemShut {NoStop}%
\bibitem [{\citenamefont {Walz-Flannigan}\ \emph {et~al.}(2004)\citenamefont
  {Walz-Flannigan}, \citenamefont {Guest}, \citenamefont {Choi},\ and\
  \citenamefont {Raithel}}]{Walz-Flannigan2004}%
  \BibitemOpen
  \bibfield  {author} {\bibinfo {author} {\bibfnamefont {A.}~\bibnamefont
  {Walz-Flannigan}}, \bibinfo {author} {\bibfnamefont {J.~R.}\ \bibnamefont
  {Guest}}, \bibinfo {author} {\bibfnamefont {J.-H.}\ \bibnamefont {Choi}}, \
  and\ \bibinfo {author} {\bibfnamefont {G.}~\bibnamefont {Raithel}},\
  }\bibfield  {title} {\enquote {\bibinfo {title} {{Cold-Rydberg-gas
  dynamics}},}\ }\href {\doibase 10.1103/PhysRevA.69.063405} {\bibfield
  {journal} {\bibinfo  {journal} {Phys. Rev. A}\ }\textbf {\bibinfo {volume}
  {69}},\ \bibinfo {pages} {063405} (\bibinfo {year} {2004})}\BibitemShut
  {NoStop}%
\bibitem [{\citenamefont {G{\"{u}}rtler}\ and\ \citenamefont {van~der
  Zande}(2004)}]{Guertler2004}%
  \BibitemOpen
  \bibfield  {author} {\bibinfo {author} {\bibfnamefont {A.}~\bibnamefont
  {G{\"{u}}rtler}}\ and\ \bibinfo {author} {\bibfnamefont {W.}~\bibnamefont
  {van~der Zande}},\ }\bibfield  {title} {\enquote {\bibinfo {title} {{l-state
  selective field ionization of rubidium Rydberg states}},}\ }\href {\doibase
  10.1016/j.physleta.2004.02.062} {\bibfield  {journal} {\bibinfo  {journal}
  {Phys. Lett. A}\ }\textbf {\bibinfo {volume} {324}},\ \bibinfo {pages} {315}
  (\bibinfo {year} {2004})}\BibitemShut {NoStop}%
\bibitem [{\citenamefont {Chibisov}\ \emph {et~al.}(2002)\citenamefont
  {Chibisov}, \citenamefont {Khuskivadze},\ and\ \citenamefont
  {Fabrikant}}]{Chibisov2002}%
  \BibitemOpen
  \bibfield  {author} {\bibinfo {author} {\bibfnamefont {M.~I.}\ \bibnamefont
  {Chibisov}}, \bibinfo {author} {\bibfnamefont {A.~A.}\ \bibnamefont
  {Khuskivadze}}, \ and\ \bibinfo {author} {\bibfnamefont {I.~I.}\ \bibnamefont
  {Fabrikant}},\ }\bibfield  {title} {\enquote {\bibinfo {title} {{Energies and
  dipole moments of long-range molecular Rydberg states}},}\ }\href {\doibase
  10.1088/0953-4075/35/10/101} {\bibfield  {journal} {\bibinfo  {journal} {J.
  Phys. B At. Mol. Opt. Phys.}\ }\textbf {\bibinfo {volume} {35}},\ \bibinfo
  {pages} {L193} (\bibinfo {year} {2002})}\BibitemShut {NoStop}%
\bibitem [{\citenamefont {Hamilton}\ \emph {et~al.}(2002)\citenamefont
  {Hamilton}, \citenamefont {Greene},\ and\ \citenamefont
  {Sadeghpour}}]{Hamilton2002}%
  \BibitemOpen
  \bibfield  {author} {\bibinfo {author} {\bibfnamefont {E.~L.}\ \bibnamefont
  {Hamilton}}, \bibinfo {author} {\bibfnamefont {C.~H.}\ \bibnamefont
  {Greene}}, \ and\ \bibinfo {author} {\bibfnamefont {H.~R.}\ \bibnamefont
  {Sadeghpour}},\ }\bibfield  {title} {\enquote {\bibinfo {title}
  {{Shape-resonance-induced long-range molecular Rydberg states}},}\ }\href
  {\doibase 10.1088/0953-4075/35/10/102} {\bibfield  {journal} {\bibinfo
  {journal} {J. Phys. B At. Mol. Opt. Phys.}\ }\textbf {\bibinfo {volume}
  {35}},\ \bibinfo {pages} {L199} (\bibinfo {year} {2002})}\BibitemShut
  {NoStop}%
\bibitem [{\citenamefont {Landau}\ and\ \citenamefont
  {Liffshitz}(1977)}]{Landau1977}%
  \BibitemOpen
  \bibfield  {author} {\bibinfo {author} {\bibfnamefont {L.~D.}\ \bibnamefont
  {Landau}}\ and\ \bibinfo {author} {\bibfnamefont {E.~M.}\ \bibnamefont
  {Liffshitz}},\ }\href {\doibase 10.1016/B978-0-08-020940-1.50001-3} {\emph
  {\bibinfo {title} {Quantum Mech.}}}\ (\bibinfo  {publisher} {Elsevier},\
  \bibinfo {year} {1977})\BibitemShut {NoStop}%
\bibitem [{\citenamefont {Clark}(1979)}]{Clark1979}%
  \BibitemOpen
  \bibfield  {author} {\bibinfo {author} {\bibfnamefont {C.~W.}\ \bibnamefont
  {Clark}},\ }\bibfield  {title} {\enquote {\bibinfo {title} {{The calculation
  of non-adiabatic transition probabilities}},}\ }\href {\doibase
  10.1016/0375-9601(79)90127-0} {\bibfield  {journal} {\bibinfo  {journal}
  {Phys. Lett. A}\ }\textbf {\bibinfo {volume} {70}},\ \bibinfo {pages} {295}
  (\bibinfo {year} {1979})}\BibitemShut {NoStop}%
\bibitem [{\citenamefont {Gaj}\ \emph {et~al.}(2014)\citenamefont {Gaj},
  \citenamefont {Krupp}, \citenamefont {Balewski}, \citenamefont {L{\"{o}}w},
  \citenamefont {Hofferberth},\ and\ \citenamefont {Pfau}}]{Gaj2014}%
  \BibitemOpen
  \bibfield  {author} {\bibinfo {author} {\bibfnamefont {A.}~\bibnamefont
  {Gaj}}, \bibinfo {author} {\bibfnamefont {A.~T.}\ \bibnamefont {Krupp}},
  \bibinfo {author} {\bibfnamefont {J.~B.}\ \bibnamefont {Balewski}}, \bibinfo
  {author} {\bibfnamefont {R.}~\bibnamefont {L{\"{o}}w}}, \bibinfo {author}
  {\bibfnamefont {S.}~\bibnamefont {Hofferberth}}, \ and\ \bibinfo {author}
  {\bibfnamefont {T.}~\bibnamefont {Pfau}},\ }\bibfield  {title} {\enquote
  {\bibinfo {title} {{From molecular spectra to a density shift in dense
  Rydberg gases}},}\ }\href {\doibase 10.1038/ncomms5546} {\bibfield  {journal}
  {\bibinfo  {journal} {Nat. Commun.}\ }\textbf {\bibinfo {volume} {5}},\
  \bibinfo {pages} {4546} (\bibinfo {year} {2014})}\BibitemShut {NoStop}%
\bibitem [{\citenamefont {Krupp}\ \emph {et~al.}(2014)\citenamefont {Krupp},
  \citenamefont {Gaj}, \citenamefont {Balewski}, \citenamefont
  {Ilzh{\"{o}}fer}, \citenamefont {Hofferberth}, \citenamefont {L{\"{o}}w},
  \citenamefont {Pfau}, \citenamefont {Kurz},\ and\ \citenamefont
  {Schmelcher}}]{Krupp2014}%
  \BibitemOpen
  \bibfield  {author} {\bibinfo {author} {\bibfnamefont {A.~T.}\ \bibnamefont
  {Krupp}}, \bibinfo {author} {\bibfnamefont {A.}~\bibnamefont {Gaj}}, \bibinfo
  {author} {\bibfnamefont {J.~B.}\ \bibnamefont {Balewski}}, \bibinfo {author}
  {\bibfnamefont {P.}~\bibnamefont {Ilzh{\"{o}}fer}}, \bibinfo {author}
  {\bibfnamefont {S.}~\bibnamefont {Hofferberth}}, \bibinfo {author}
  {\bibfnamefont {R.}~\bibnamefont {L{\"{o}}w}}, \bibinfo {author}
  {\bibfnamefont {T.}~\bibnamefont {Pfau}}, \bibinfo {author} {\bibfnamefont
  {M.}~\bibnamefont {Kurz}}, \ and\ \bibinfo {author} {\bibfnamefont
  {P.}~\bibnamefont {Schmelcher}},\ }\bibfield  {title} {\enquote {\bibinfo
  {title} {{Alignment of D-state Rydberg molecules}},}\ }\href {\doibase
  10.1103/PhysRevLett.112.143008} {\bibfield  {journal} {\bibinfo  {journal}
  {Phys. Rev. Lett.}\ }\textbf {\bibinfo {volume} {112}},\ \bibinfo {pages}
  {143008} (\bibinfo {year} {2014})}\BibitemShut {NoStop}%
\bibitem [{\citenamefont {Holmgren}\ \emph {et~al.}(2010)\citenamefont
  {Holmgren}, \citenamefont {Revelle}, \citenamefont {Lonij},\ and\
  \citenamefont {Cronin}}]{Holmgren2010}%
  \BibitemOpen
  \bibfield  {author} {\bibinfo {author} {\bibfnamefont {W.~F.}\ \bibnamefont
  {Holmgren}}, \bibinfo {author} {\bibfnamefont {M.~C.}\ \bibnamefont
  {Revelle}}, \bibinfo {author} {\bibfnamefont {V.~P.~A.}\ \bibnamefont
  {Lonij}}, \ and\ \bibinfo {author} {\bibfnamefont {A.~D.}\ \bibnamefont
  {Cronin}},\ }\bibfield  {title} {\enquote {\bibinfo {title} {{Absolute and
  ratio measurements of the polarizability of Na, K, and Rb with an atom
  interferometer}},}\ }\href {\doibase 10.1103/PhysRevA.81.053607} {\bibfield
  {journal} {\bibinfo  {journal} {Phys. Rev. A}\ }\textbf {\bibinfo {volume}
  {81}},\ \bibinfo {pages} {053607} (\bibinfo {year} {2010})}\BibitemShut
  {NoStop}%
\bibitem [{\citenamefont {Bendkowsky}\ \emph {et~al.}(2010)\citenamefont
  {Bendkowsky}, \citenamefont {Butscher}, \citenamefont {Nipper}, \citenamefont
  {Balewski}, \citenamefont {Shaffer}, \citenamefont {L{\"{o}}w}, \citenamefont
  {Pfau}, \citenamefont {Li}, \citenamefont {Stanojevic}, \citenamefont
  {Pohl},\ and\ \citenamefont {Rost}}]{Bendkowsky2010}%
  \BibitemOpen
  \bibfield  {author} {\bibinfo {author} {\bibfnamefont {V.}~\bibnamefont
  {Bendkowsky}}, \bibinfo {author} {\bibfnamefont {B.}~\bibnamefont
  {Butscher}}, \bibinfo {author} {\bibfnamefont {J.}~\bibnamefont {Nipper}},
  \bibinfo {author} {\bibfnamefont {J.~B.}\ \bibnamefont {Balewski}}, \bibinfo
  {author} {\bibfnamefont {J.~P.}\ \bibnamefont {Shaffer}}, \bibinfo {author}
  {\bibfnamefont {R.}~\bibnamefont {L{\"{o}}w}}, \bibinfo {author}
  {\bibfnamefont {T.}~\bibnamefont {Pfau}}, \bibinfo {author} {\bibfnamefont
  {W.}~\bibnamefont {Li}}, \bibinfo {author} {\bibfnamefont {J.}~\bibnamefont
  {Stanojevic}}, \bibinfo {author} {\bibfnamefont {T.}~\bibnamefont {Pohl}}, \
  and\ \bibinfo {author} {\bibfnamefont {J.~M.}\ \bibnamefont {Rost}},\
  }\bibfield  {title} {\enquote {\bibinfo {title} {{Rydberg trimers and excited
  dimers bound by internal quantum reflection}},}\ }\href {\doibase
  10.1103/PhysRevLett.105.163201} {\bibfield  {journal} {\bibinfo  {journal}
  {Phys. Rev. Lett.}\ }\textbf {\bibinfo {volume} {105}},\ \bibinfo {pages}
  {163201} (\bibinfo {year} {2010})}\BibitemShut {NoStop}%
\bibitem [{\citenamefont {Eiles}\ \emph {et~al.}(2016)\citenamefont {Eiles},
  \citenamefont {Perez-Rios}, \citenamefont {Robicheaux},\ and\ \citenamefont
  {Greene}}]{Eiles2016}%
  \BibitemOpen
  \bibfield  {author} {\bibinfo {author} {\bibfnamefont {M.~T.}\ \bibnamefont
  {Eiles}}, \bibinfo {author} {\bibfnamefont {J.}~\bibnamefont {Perez-Rios}},
  \bibinfo {author} {\bibfnamefont {F.}~\bibnamefont {Robicheaux}}, \ and\
  \bibinfo {author} {\bibfnamefont {C.~H.}\ \bibnamefont {Greene}},\ }\bibfield
   {title} {\enquote {\bibinfo {title} {{Ultracold molecular Rydberg physics in
  a high density environment}},}\ }\href {http://arxiv.org/abs/1601.06881}
  {\bibfield  {journal} {\bibinfo  {journal} {arXiv}\ } (\bibinfo {year}
  {2016})},\ \Eprint {http://arxiv.org/abs/1601.06881} {arXiv:1601.06881}
  \BibitemShut {NoStop}%
\bibitem [{\citenamefont {Saffman}\ \emph {et~al.}(2010)\citenamefont
  {Saffman}, \citenamefont {Walker},\ and\ \citenamefont
  {M{\o}lmer}}]{Saffman2010}%
  \BibitemOpen
  \bibfield  {author} {\bibinfo {author} {\bibfnamefont {M.}~\bibnamefont
  {Saffman}}, \bibinfo {author} {\bibfnamefont {T.~G.}\ \bibnamefont {Walker}},
  \ and\ \bibinfo {author} {\bibfnamefont {K.}~\bibnamefont {M{\o}lmer}},\
  }\bibfield  {title} {\enquote {\bibinfo {title} {{Quantum information with
  Rydberg atoms}},}\ }\href {\doibase 10.1103/RevModPhys.82.2313} {\bibfield
  {journal} {\bibinfo  {journal} {Rev. Mod. Phys.}\ }\textbf {\bibinfo {volume}
  {82}},\ \bibinfo {pages} {2313} (\bibinfo {year} {2010})}\BibitemShut
  {NoStop}%
\bibitem [{\citenamefont {Faoro}\ \emph {et~al.}(2016)\citenamefont {Faoro},
  \citenamefont {Simonelli}, \citenamefont {Archimi}, \citenamefont {Masella},
  \citenamefont {Valado}, \citenamefont {Arimondo}, \citenamefont {Mannella},
  \citenamefont {Ciampini},\ and\ \citenamefont {Morsch}}]{Faoro2016}%
  \BibitemOpen
  \bibfield  {author} {\bibinfo {author} {\bibfnamefont {R.}~\bibnamefont
  {Faoro}}, \bibinfo {author} {\bibfnamefont {C.}~\bibnamefont {Simonelli}},
  \bibinfo {author} {\bibfnamefont {M.}~\bibnamefont {Archimi}}, \bibinfo
  {author} {\bibfnamefont {G.}~\bibnamefont {Masella}}, \bibinfo {author}
  {\bibfnamefont {M.~M.}\ \bibnamefont {Valado}}, \bibinfo {author}
  {\bibfnamefont {E.}~\bibnamefont {Arimondo}}, \bibinfo {author}
  {\bibfnamefont {R.}~\bibnamefont {Mannella}}, \bibinfo {author}
  {\bibfnamefont {D.}~\bibnamefont {Ciampini}}, \ and\ \bibinfo {author}
  {\bibfnamefont {O.}~\bibnamefont {Morsch}},\ }\bibfield  {title} {\enquote
  {\bibinfo {title} {{van der Waals explosion of cold Rydberg clusters}},}\
  }\href {\doibase 10.1103/PhysRevA.93.030701} {\bibfield  {journal} {\bibinfo
  {journal} {Phys. Rev. A}\ }\textbf {\bibinfo {volume} {93}},\ \bibinfo
  {pages} {030701} (\bibinfo {year} {2016})}\BibitemShut {NoStop}%
\end{thebibliography}
\end{document}